\documentclass[%
 reprint,
superscriptaddress,
showpacs,
preprintnumbers,
 amsmath,amssymb,
pra,
longbibliography
]{revtex4-1}

\usepackage[dvipdfmx]{graphicx}

\usepackage{physics}
\usepackage{tikz}
\usetikzlibrary{quantikz2}

\usepackage{dcolumn}
\usepackage{bm}
\usepackage{setspace}	
\usepackage{color}
\usepackage{amsmath}	
\usepackage{orcidlink}

\usepackage{upgreek}

\DeclareGraphicsExtensions{.eps}
\graphicspath{{figs/}}

\def\ii{{\mathrm{i}}}

\def\ee{{\mathrm{e}}}

\def\Hc{\mathrm{H.c.}} 

\def\no{{\nonumber}} 

\def\bra#1{\langle #1|}

\def\ket#1{|#1\rangle}

\def\bracket#1{\langle #1 \rangle}
\def\bracketi#1#2{\langle #1 | #2 \rangle}
\def\bracketii#1#2#3{\langle #1 | #2| #3\rangle}

\def\sub#1{_\mathrm{#1}} 
\def\sur#1{^\mathrm{#1}} 

\def\tr{\mathrm{tr}}

\begin{document}


\title{
High-yield integration design of fixed-frequency superconducting qubit systems\\
using siZZle-CZ gates
}


\author{Kazuhisa Ogawa\,\orcidlink{0000-0003-3096-6669}}
\email{k-ogawa.qiqb@osaka-u.ac.jp}
\affiliation{%
Center for Quantum Information and Quantum Biology (QIQB), The University of Osaka, Toyonaka, Osaka 560-0043, Japan
}%

\author{Yutaka Tabuchi\,\orcidlink{0000-0003-2512-1856}}
\affiliation{%
RIKEN Center for Quantum Computing, Wako, Saitama 351-0198, Japan
}%

\author{Makoto Negoro\,\orcidlink{0000-0003-3482-3198}}
\affiliation{%
Center for Quantum Information and Quantum Biology (QIQB), The University of Osaka, Toyonaka, Osaka 560-0043, Japan
}%
\affiliation{
Graduate School of Engineering Science, The University of Osaka, Toyonaka, Osaka 560-0043, Japan
}
\affiliation{
QuEL,~Inc., Daiwaunyu Building 3F, 2-9-2 Owadamachi, Hachioji, Tokyo 192-0045, Japan
}

\date{\today}

\begin{abstract}
Fixed-frequency transmon qubits, characterized by simple architectures and long coherence times, are promising platforms for large-scale quantum computing.
However, the rapidly increasing frequency collisions, which directly reduce the fabrication yield, hinder scaling, especially in cross-resonance (CR) gate-based architectures, wherein the restricted drive frequency severely limits the available design space.
We investigate the Stark-induced ZZ by level excursions (siZZle) gate, which relaxes this limitation by allowing arbitrary drive-frequency choices.
Extensive numerical analyses across a broad parameter range---including the far-detuned regime that has received negligible prior attention---reveal wide operating windows that support controlled-Z (CZ) fidelities $>99.6\%$.
Leveraging these windows, we design lattice architectures containing $>1000$ qubits, showing that even under $0.25\%$ fabrication-induced frequency dispersion, the zero-collision yields in square and heavy-hexagonal lattices reach $80\%$ and $100\%$, respectively.
Thus, the siZZle-CZ gate is a scalable and collision-robust alternative to the CR gate, offering a viable route toward high-yield fixed-frequency transmon quantum processors.
\end{abstract}

\maketitle


\begin{spacing}{1.}

\section{Introduction}


The realization of large-scale quantum computers is a crucial challenge in the field of quantum information processing. 
Architectures based on superconducting qubits have emerged as one of the most promising technologies among the various physical platforms currently under active investigation \cite{nature2024quantum, rqkg-dw31, kim2023evidence, krinner2022realizing}. 
Recent advances have led to substantial improvements in both the coherence times of superconducting qubits and the performance of two-qubit (2Q) gates, paving the way toward scalable quantum processors. 
In particular, fixed-frequency transmon qubits \cite{kim2023evidence, acharya2025integration, hashim2025quasiprobabilistic, PRXQuantum.6.020345, PhysRevX.11.041039, spring2022high}, which are robust against charge and flux noise and require only a single control line per qubit, reduce potential noise injection from control wiring. 
Owing to their relatively long coherence times and simple hardware configuration, fixed-frequency transmon qubits are considered as viable candidates for large-scale quantum computing.
In fixed-frequency transmon architectures, the cross-resonance (CR) gate is the most commonly employed as the native 2Q gate \cite{PhysRevB.81.134507,PhysRevLett.107.080502, PhysRevA.93.060302, PhysRevApplied.12.064013,PhysRevA.102.042605,PRXQuantum.1.020318}. 
Upon applying a microwave drive (at the resonance frequency of the target qubit) to the control qubit, the CR gate induces effective ZX interactions, enabling the implementation of high-fidelity CNOT gates approaching the coherence limit, with gate durations of the order of several hundred nanoseconds.

\begin{figure*}[t]
\centering
\includegraphics[width=18cm]{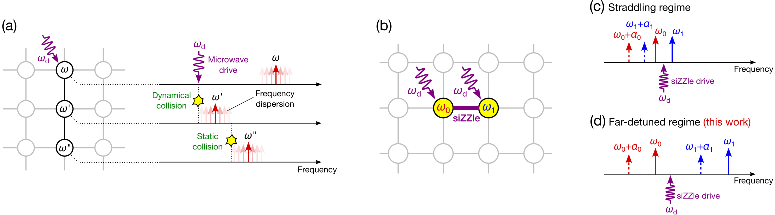}
\caption{
(a) Schematic of frequency collisions in integrated arrays of fixed-frequency transmon qubits.
In practice, the qubit resonance frequencies exhibit stochastic dispersion from their design values owing to fabrication imperfections, probabilistically leading to static collisions, in which resonance transition frequencies of nearby qubits become closely spaced, as well as to dynamical collisions, in which microwave-drive crosstalk induces excitations of neighboring qubits.
(b) siZZle drive applied to two qubits in an integrated qubit array.
An effective ZZ interaction is induced by simultaneously driving two neighboring qubits with a microwave tone at the same frequency $\omega_{\mathrm{d}}$.
(c), (d) Relative configuration of the qubit resonance frequencies ($\omega_0$ and $\omega_1$) and the siZZle drive frequency ($\omega_{\mathrm{d}}$).
The qubit detuning $\varDelta_{10} = \omega_1 - \omega_0$ can be classified into the straddling regime, where $|\varDelta_{10}|$ is smaller than the anharmonicities $|\alpha_0|$ and $|\alpha_1|$, and the far-detuned regime, where $|\varDelta_{10}|$ exceeds the anharmonicities.
The siZZle drive frequency $\omega_{\mathrm{d}}$ can be chosen to a certain extent, and the resulting strength of the ZZ interaction depends on this choice.
}
\label{fig:abst}
\end{figure*}

A major challenge in scaling up fixed-frequency transmon-based systems is the problem of frequency collisions \cite{hertzberg2021laser,morvan2022optimizing,PhysRevResearch.5.043001,zhang2022high,PhysRevA.111.012619,smith2022scaling}. 
As illustrated in Fig.~\ref{fig:abst}(a), frequency collisions occur when the transition frequencies of neighboring qubits or the drive frequencies used for gate operations are close to other transition frequencies, leading to unintended excitations and consequently enhanced gate errors. 
To mitigate frequency collisions, the lattice geometry and frequency allocation are typically carefully designed in integrated fixed-frequency transmon systems. 
However, in practice, the resonance frequencies of superconducting qubits exhibit stochastic dispersion, which is primarily attributed to fabrication imperfections in Josephson junctions. 
Consequently, the zero-collision yield---the probability that no frequency collision occurs---decreases exponentially with increasing system size.
Post-fabrication techniques such as laser annealing reportedly minimize the qubit frequency dispersion to 0.25\% \cite{hertzberg2021laser}. 
However, even under this assumption, the estimated zero-collision yield for a 127-qubit heavy-hexagonal lattice is only 8\%, which drops to below 0.1\% for systems comprising approximately 1000 qubits \cite{hertzberg2021laser}. 
Subsequent studies have explored optimized frequency allocation strategies, demonstrating that to achieve a 10\% yield in square lattices with more than 1000 qubits, the qubit frequency dispersion must be reduced to approximately 0.14\% \cite{PhysRevA.111.012619}.

In this study, to further improve the zero-collision yield, we focus on a controlled-Z (CZ) gate based on the Stark-induced ZZ by level excursions (siZZle) mechanism as an alternative 2Q gate to the CR gate \cite{wei2022hamiltonian,wei2024native,mitchell2021hardware,nguyen2024programmable,alghadeer2025low,cao2025automating,n6gs-zlhc}. 
As illustrated in Fig.~\ref{fig:abst}(b), siZZle induces an effective ZZ interaction by simultaneously applying off-resonant microwave drives at the same frequency $\omega_{\mathrm{d}}$ to the two qubits intended for interaction. 
This interaction can be exploited to construct a CZ gate, referred to as the siZZle-CZ gate (see Sec.~\ref{sec:sizzle} for theoretical details of siZZle). 
Proof-of-principle experiments of the siZZle-CZ gate in small-scale systems have reported gate fidelities exceeding 99.4\% \cite{wei2022hamiltonian,mitchell2021hardware}.
In CR gates, the drive frequency is set to match the resonance frequency of the target qubit, thereby increasing the likelihood of frequency collisions with the resonance frequencies of other qubits.
By contrast, in siZZle-CZ gates, the drive frequency can be selected with a degree of flexibility, allowing for the effective avoidance of collisions associated with the drive frequency; this capability enables zero-collision yield enhancement.
For example, with a qubit frequency dispersion of 0.25\% and target 2Q gate fidelity of approximately 99\%, a 1000-qubit integrated chip configured in a heavy-hexagonal lattice can achieve a zero-collision yield of 10\%.\cite{morvan2022optimizing}.

To realize fault-tolerant quantum computation, an increasing number of qubits with higher 2Q gate fidelities and high connectivity on lattice architectures, such as square lattices, must be integrated while maintaining a high zero-collision yield. 
To address this requirement, in this study, we numerically analyze the optimal conditions for qubit resonance and drive frequencies that enable high-fidelity siZZle-CZ gates and evaluate the achievable zero-collision yield in integrated quantum systems employing siZZle-CZ gates as native 2Q gates.
In particular, we analyze the qubit detuning $\varDelta_{10}$ between the two qubits participating in a siZZle-CZ gate over a broader frequency-detuning range; this range encompasses the straddling regime [Fig.~\ref{fig:abst}(c)], which has been widely adopted for transmon qubits in previous studies, as well as the far-detuned regime [Fig.~\ref{fig:abst}(d)], defined by $|\varDelta_{10}| > |\alpha_0|, |\alpha_1|$ (where $\alpha_0$ and $\alpha_1$ denote the anharmonicities of the respective qubits).
Frequency configurations in the far-detuned regime have been reportedly employed in several transmon-based systems \cite{PRXQuantum.6.020345, PhysRevApplied.14.044039}. 
Although this regime typically requires microwave drives with relatively large amplitudes and broad bandwidths, it is less susceptible to frequency collisions caused by fabrication-induced dispersions in qubit frequencies.
Consequently, by combining qubit frequency allocations based on the far-detuned regime with siZZle-CZ gates, which allow flexibility in the choice of drive frequencies, frequency collisions can be effectively suppressed, and an improved fabrication yield is expected in integrated qubit systems. 
In this study, first, we numerically explore a wide frequency parameter space---including the far-detuned regime---by sweeping the qubit detuning and siZZle drive frequency to identify parameter regions in which high-fidelity siZZle-CZ gates can be realized. 
Then, based on these results, we design optimal frequency allocations for integrated qubit systems employing siZZle-CZ gates and numerically evaluate the zero-collision yield of integrated chips while accounting for the fabrication-induced qubit frequency dispersion. The results demonstrate that under these optimized conditions, the zero-collision yield can approach nearly 100\% even for large-scale integrated chips with more than 1000 qubits.

\section{Results}\label{sec:2}

\subsection{Investigation of a valid parameter range for siZZle-CZ gates}\label{sec:sizzle_calc}

\begin{figure*}[t]
\centering
\includegraphics[width=18cm]{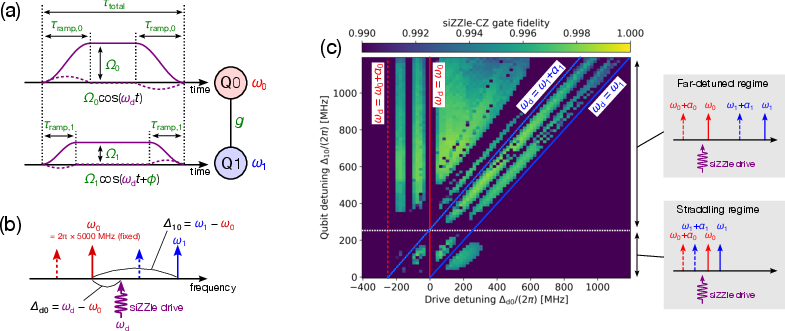}
\caption{
(a) Parameters used in the numerical simulations of the siZZle dynamics (see the main text and Table \ref{tab:1} for detailed numerical values).
(b) Arrangement of the resonance frequencies of the two qubits and siZZle drive frequency.
The swept parameters are the drive detuning relative to the g-e transition of Q0, $\varDelta_{\rm d0} := \omega_{\rm d} - \omega_0$, 
and the qubit--qubit detuning, $\varDelta_{10} := \omega_1 - \omega_0$,
where the g-e resonance frequency of Q0 is fixed to $\omega_0 = 2\pi \times 5000~{\rm MHz}$.
In the straddling regime, either $\omega_1 + \alpha_1 < \omega_0 < \omega_1$ or $\omega_0 + \alpha_0 < \omega_1 < \omega_0$ is satisfied, whereas in the far-detuned regime either $\omega_0 < \omega_1 + \alpha_1$ or $\omega_1 < \omega_0 + \alpha_0$ holds.
(c) Numerically calculated fidelity of the siZZle-CZ gate as a function of the qubit detuning $\varDelta_{10}$ and drive detuning $\varDelta_{\rm d0}$.
The red solid and dashed lines indicate the g-e and e-f resonance frequencies of Q0, respectively, and the blue solid and dashed lines denote the corresponding transitions of Q1.
}
\label{fig:1}
\end{figure*}

\begin{table}[t]
    \centering
    \begin{tabular}{ccc}
        \hline\hline
         & Qubit 0 & Qubit 1\\
        \hline
        $\omega_{i} / (2\pi)$ & $\qquad 5000\,{\rm MHz}\qquad$ & $5000\,{\rm MHz} + \varDelta_{10}/(2\pi)$ \\
        $\varDelta_{10}/(2\pi)$ & \multicolumn{2}{c}{Sweep parameter} \\
        $\alpha_i / (2\pi)$ & $-250\,{\rm MHz}$ & $-0.05\times\omega_1/(2\pi)$ \\
        $g/(2\pi)$ & \multicolumn{2}{c}{Pre-defined function of $\varDelta_{10}$} \\
        $\varDelta_{\rm d0}/ (2\pi)$ & \multicolumn{2}{c}{Sweep parameter} \\
        $\tau_{{\rm ramp},i}$ & Optimized & Optimized \\
        $r_{{\rm iq},i}$ & Optimized & Optimized \\
        $\varOmega_{i}/ (2\pi)$ & Optimized & Optimized \\
        $\tau_{\rm total}$ & \multicolumn{2}{c}{Optimized} \\
        \hline\hline
    \end{tabular}
    \caption{
    Parameters used in the numerical simulations.
    $\omega_i$: g-e transition frequency of qubit Q-$i$ ($i=0,1$);
    $\varDelta_{10} = \omega_1 - \omega_0$: qubit detuning;
    $\alpha_i$: anharmonicity of qubit Q-$i$;
    $g$: qubit--qubit coupling strength;
    $\varDelta_{\rm d0} = \omega_{\rm d} - \omega_0$: siZZle drive detuning;
    $\tau_{{\rm ramp},i}$: ramp time of the drive pulse;
    $r_{{\rm iq},i}$: ratio between the in-phase and quadrature components of the DRAG waveform in the ramp segment of the drive pulse;
    $\varOmega_i$: maximum amplitude of the drive pulse;
    $\tau_{{\rm total}}$: total duration of the drive pulse.
    The parameters $\varDelta_{10}/(2\pi)$ and $\varDelta_{\rm d0}/(2\pi)$ are swept in the ranges $0$\,--$1200$\, and $-400$\,--$1200$\,MHz, respectively.
    The coupling strength $g$ is defined \textit{a priori} as a function of $\varDelta_{10}$ such that the gate errors induced by the frequency collisions with neighboring idle qubits are sufficiently suppressed in the subsequent yield analysis (see Sec.\ref{sec:g} for details).
    The pulse parameters $\tau_{{\rm ramp},i}$, $r_{{\rm iq},i}$, $\varOmega_i$, and $\tau_{{\rm total}}$ are optimized for each set of swept parameters ($\varDelta_{10}$, $\varDelta_{\rm d0}$) to maximize the fidelity of the siZZle-CZ gate (see Sec.\ref{sec:sizzle_params} for details).    
     }
    \label{tab:1}
\end{table}


We systematically evaluate the ranges of the qubit and drive detunings suitable for realizing high-fidelity siZZle-CZ gates.
In particular, we identify optimal parameter regimes for qubit detuning in the far-detuned regime, which has not been sufficiently explored in previous studies.
In this analysis, we simulate the time evolution of the 2Q system under siZZle driving via numerical calculations using QuTiP \cite{lambert2024qutip, qutip}.
The gate fidelity is evaluated by numerically calibrating the siZZle-CZ gate and 2Q quantum process tomography; detailed procedures for the siZZle-CZ calibration and 2Q process tomography are described in Secs.~\ref{sec:sizzle_calib} and \ref{sec:qpt}, respectively.


Figure~\ref{fig:1}(a) depicts the system considered in this study.
The two transmon qubits, Q0 and Q1, with g-e transition frequencies $\omega_i$ and anharmonicities $\alpha_i$ ($i=0,1$), are modeled as four-level anharmonic systems.
The energy and phase relaxation times ($T_1$ and $T_2^*$, respectively) are both set as $600~\mu{\rm s}$ \cite{abughanem2025ibm}.
Q0 and Q1 are coupled with a coupling strength $g$, and siZZle microwave pulses, with a common drive frequency of $\omega_{\rm d}$, are applied simultaneously to both qubits.
The drive pulses are assumed to have the derivative-removal-by-adiabatic-gate (DRAG) envelopes \cite{PhysRevLett.103.110501} with cosine-shaped rise and fall segments; the ratio between the in-phase and quadrature components is denoted by $r_{{\rm iq},i}$, and the pulse amplitude, ramp time, and total duration are denoted by $\varOmega_i$, $\tau_{{\rm ramp},i}$, and $\tau_{\rm total}$, respectively.
The numerical values of all the parameters are summarized in Table\ref{tab:1}.


As illustrated in Fig.~\ref{fig:1}(b), in this study, we fix the g-e transition frequency of Q0 to $\omega_0 = 2\pi \times 5000~{\rm MHz}$ and utilize the following parameters for sweeping:
\begin{align}
\varDelta_{10} := \omega_1 - \omega_0,\quad
\varDelta_{\rm d0} := \omega_{\rm d} - \omega_0 .
\end{align}
In particular, the qubit detuning $\varDelta_{10}$ is swept within the straddling regime as well as is extended into the far-detuned regime.
The remaining parameters, indicated in green in Fig.\ref{fig:1}(a), are determined for each pair of $\varDelta_{\rm d0}$ and $\varDelta_{10}$ according to the following criteria:
\begin{itemize}
\item qubit--qubit coupling strength $g$:
Increasing $g$ shortens the pulse duration required for the CZ gate and thereby improves the gate fidelity; however, it also increases the probability of frequency collisions, which reduces the chip yield.
To account for this trade-off, we define \textit{a priori} a function $g(E_{\rm idle}$, $\varDelta_{10}$) such that the gate error induced by neighboring idle qubits during a fixed idle time is limited to a sufficiently small constant $E_{\rm idle}$, ensuring a negligible frequency-collision probability in the subsequent yield analysis.
This function is used throughout the simulations (see Sec.\ref{sec:g} for details).
\item Ratio between in-phase and quadrature components of the DRAG waveform $r_{{\rm iq},i}$:
For each value of drive detuning $\varDelta_{{\rm d}i}$ and ramp time $\tau_{{\rm ramp},i}$, the value of $r_{{\rm iq},i}$ is selected to minimize the probability of nonadiabatic transitions (see Sec.\ref{sec:rcft_drag} for details).
\item Ramp time $\tau_{{\rm ramp},i}$ and drive amplitude $\varOmega_i$:
These parameters are selected to maximize the pulse area while maintaining the nonadiabatic transition probability below $10^{-4}$.
The maximum values of $\tau_{{\rm ramp},i}$ and $\varOmega_i$ are set as $100\,{\rm ns}$ and $200\,{\rm MHz}$, respectively (see Sec.\ref{sec:sizzle_params} for details).
\item Total pulse duration $\tau_{\rm total}$:
This parameter is calibrated such that the unitary evolution generated by the effective ZZ interaction realizes a CZ gate.
\item Relative phase $\phi$ between the two siZZle pulses:
The magnitude of the siZZle-induced ZZ interaction is maximized when $\phi = 0$ or $\pi$, whereas its sign is inverted between these two cases.
Depending on the values of $\varDelta_{10}$ and $\varDelta_{\rm d0}$, the sign of the siZZle-induced ZZ interaction varies, leading to parameter regions where it interferes constructively or destructively with the statistic ZZ interaction.
In each region, we select the value of $\phi$ such that the sum of the siZZle-induced and statistic ZZ interactions interferes constructively (see Sec.~\ref{sec:parametric} for details).
\end{itemize}


Figure~\ref{fig:1}(c) presents a colormap showing the maximum fidelity of the siZZle-CZ gate as a function of the qubit detuning $\varDelta_{10}$ and drive detuning $\varDelta_{\rm d0}$.
The lower region of the plot, corresponding to $\varDelta_{10} < -\alpha_1$, represents the straddling regime, in which frequency conditions enabling high-fidelity operation have been reported in previous studies~\cite{wei2022hamiltonian, mitchell2021hardware}.
In contrast, the upper region, where $\varDelta_{10} > -\alpha_1$, corresponds to the far-detuned regime, in which fidelities exceeding 99.6\% are achievable over a substantially wider range of both $\varDelta_{10}$ and $\varDelta_{\rm d0}$ than in the straddling regime.
The wide tolerance in $\varDelta_{10}$ implies that high-fidelity siZZle-CZ gates can be realized robustly against fabrication-induced dispersion in the qubit resonance frequencies.
Moreover, the broad allowable range of $\varDelta_{\rm d0}$ indicates that in contrast to CR-based schemes, the siZZle drive frequency can be adopted with greater flexibility to avoid frequency collision.
These results suggest that frequency-allocation strategies based on the far-detuned regime are suitable for mitigating frequency collisions and improving chip fabrication yield.
Finally, the line-shaped regions of reduced fidelity, observed in the colormap, originate from spectral collisions among various 2Q transitions or from those between these transitions and the siZZle drive frequency.
The dominant collision conditions, including higher-order transitions, are summarized in Sec.~\ref{sec:parametric}.

\subsection{Evaluation of qubit chip yield}\label{sec:yield}

\begin{figure*}[t]		
\centering
\includegraphics[width=18cm]{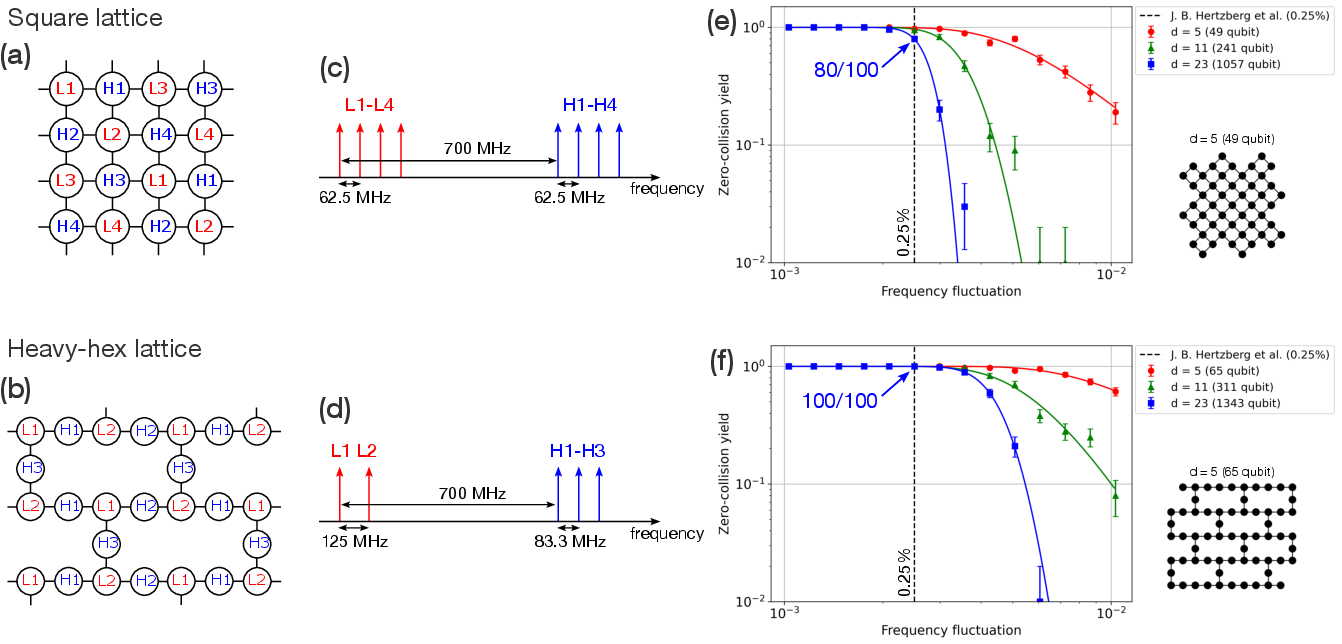}
\caption{
(a), (b) Square and heavy-hexagonal lattices considered in this study.
Qubits labeled H1--H4 and L1--L4 denote relatively high- and low-frequency qubits, respectively, which are arranged in a checkerboard pattern such that the frequency detuning between nearest-neighbor (NN) qubits lies in the far-detuned regime.
(c), (d) Specific assignments of the H1--H4 and L1--L4 qubit frequencies for the square and heavy-hexagonal lattices, respectively.
In both cases, the lowest frequency L1 is fixed to $5000\,\mathrm{MHz}$.
(e), (f) Numerically calculated zero-collision yield (vertical axis) as a function of the ratio of fabrication-induced frequency dispersion to the designed qubit frequency (horizontal axis), for the square and heavy-hexagonal lattices, respectively.
The dashed line indicates the minimum achievable frequency dispersion ratio of $0.25\%$ reported in a previous study~\cite{hertzberg2021laser}.
The red circles, green triangles, and blue squares represent yield evaluation results for square and heavy-hexagonal lattices with code distances $d=5, 11,$ and $23$, respectively.
The solid curves represent the fitting results obtained using an error-function-based model.
At the minimum achievable frequency dispersion ratio of $0.25\%$, the zero-collision yields for square and heavy-hexagonal lattices, with a code distance $d=23$, reach $80\%$ and $100\%$, respectively indicating that collision-free operation can be achieved with a realistic number of chip fabrication runs.
}
\label{fig:3}
\end{figure*}

Based on the discussion in Sec.\ref{sec:sizzle_calc}, we evaluate the presence of frequency collisions in transmon qubit arrays, whose frequency allocation is designed in the far-detuned regime, and numerically estimate the zero-collision yield for various lattice geometries and system sizes.
As qubit lattice structures, we consider the square and heavy-hexagonal lattices shown in Figs.~\ref{fig:3}(a) and \ref{fig:3}(b), respectively.
The system sizes correspond to code distances $d=5, 11,$ and $23$ for the surface \cite{nature2024quantum} and heavy-hexagonal codes \cite{sundaresan2023demonstrating}.
To assess robustness against fabrication-induced variations in qubit frequencies, we sweep the magnitude of stochastic dispersion from the designed qubit frequencies and numerically evaluate the resulting zero-collision yield under each condition.

We next describe the definition of frequency collisions used in this evaluation.
In a previous study \cite{hertzberg2021laser}, frequency collisions were defined by imposing fixed thresholds on the detunings between the transmon transition and drive frequencies \cite{morvan2022optimizing, PhysRevA.111.012619}.
However, such criteria depend strongly on parameters including qubit frequencies, qubit--qubit coupling strengths, drive amplitudes, and target gate fidelities, and therefore cannot be uniformly applied across different qubit designs.
In the present study, for each qubit lattice and frequency allocation, we explicitly compute the errors incurred by each qubit, via numerical simulations of quantum dynamics performed with QuTiP, and determine the presence of frequency collisions based on these error estimates.
 
Specifically, a frequency allocation is defined to be collision free with target error $E_{\star}$ if all the following conditions are satisfied:
\begin{enumerate}
\item For all qubits in the lattice, the idling error within an average siZZle-CZ gate duration of $700\,\mathrm{ns}$ is below $E_{\star}$.
The total error is obtained by converting the following three processes into depolarizing channels and summing them:
(i) single-qubit decoherence,
(ii) static collisions originating from nearest-neighbor (NN) and next-nearest-neighbor (NNN) spectator qubits,
and (iii) dynamic collisions induced by spectator qubits during siZZle-CZ driving.
\item For all NN qubit pairs, the siZZle-CZ gate error is below $E_{\star}$.
Here, we use the results presented in Sec.\ref{sec:sizzle_calc} as a function of the qubit ($\varDelta_{10}$) and drive detuning ($\varDelta_{\rm d0}$).
The drive frequency is selected such that it maximizes the siZZle-CZ gate fidelity while remaining within a regime where it does not significantly increase the single-qubit idling error of surrounding spectator qubits.
\end{enumerate}
If either of these conditions is violated, then the corresponding frequency allocation is classified as ``colliding'' (see Sec.\ref{sec:collision} for further details of the collision-evaluation procedure).

In this study, we set the target error as $E_{\star}=0.6\%$.
In the previous study \cite{hertzberg2021laser}, the noncollision condition was defined assuming CR-gate errors of approximately $1\%$; similarly, in subsequent studies, conducted using the same criterion~\cite{morvan2022optimizing, PhysRevA.111.012619}, comparable 2Q gate errors were assumed.
Compared with the criteria adopted in these studies, our noncollision criterion is more stringent as it enforces a target error of $E_{\star}=0.6\%$ as an explicit accuracy guarantee and explicitly accounts for both 2Q gate errors and single-qubit idling errors.

The designed frequency allocations for the square and heavy-hexagonal lattices are shown in Figs.~\ref{fig:3}(c) and \ref{fig:3}(d), respectively.
In the far-detuned regime, the detuning between adjacent qubits must exceed the magnitude of the anharmonicity; accordingly, low- and high-frequency qubits are arranged alternately in a checkerboard pattern.
To further suppress frequency collisions among NNN qubits, additional frequency offsets are introduced within both the low- and high-frequency qubit groups.
The detunings between the low- and high-frequency groups, as well as those within each group, are optimized by sweeping these parameters and numerically evaluating the resulting zero-collision yield to select the detuning conditions that maximize the yield.

The numerical results for the zero-collision yield of the square and heavy-hexagonal lattices are presented in Figs.~\ref{fig:3}(e) and \ref{fig:3}(f), respectively.
The horizontal axis represents the ratio $\Delta f$ of fabrication-induced qubit frequency dispersion to the designed qubit resonance frequency, and the dashed line indicates the minimum achievable frequency dispersion $0.25\%$ reported in Ref.~\cite{hertzberg2021laser}.
The red circles, green triangles, and blue squares correspond to system sizes associated with code distances $d=5, 11,$ and $23$, respectively, for the surface and heavy-hexagonal codes.
For the surface code, these code distances ($d=5$, $11$, and $23$) correspond to qubit numbers of 49, 241, and 1057, respectively, whereas for the heavy-hexagonal code, they correspond to 65, 311, and 1343 qubits, respectively.
The solid curves are obtained from fitting with the function
$F_{\rm fit}(\Delta f) = \mathrm{erf}(a/\Delta f)^b$,
where $\mathrm{erf}(x)$ denotes the error function, and $a$ and $b$ are fitting parameters.
These results show that for square lattices with a code distance $d=23$ (exceeding $1000$ qubits), a high zero-collision yield of $80\%$ can be achieved for a frequency variation of $0.25\%$, whereas under the same conditions, the heavy-hexagonal lattice achieves a zero-collision yield of $100\%$.
This result indicates that if the siZZle-CZ gate can be used in the far-detuned regime as the 2Q gate, then frequency allocation designs can realistically achieve system sizes exceeding $1000$ qubits with practical fabrication yields for both lattice geometries.
However, notably, the error-correction thresholds differ between the surface and heavy-hexagonal codes.
The target error $E_{\star}=0.6\%$ adopted in this study lies below the threshold for the surface code, whereas it remains above the threshold for the heavy-hexagonal code \cite{Benito2025comparativestudyof}.

\section{Discussion}

\begin{figure}[t]		
\centering
\includegraphics[width=8cm]{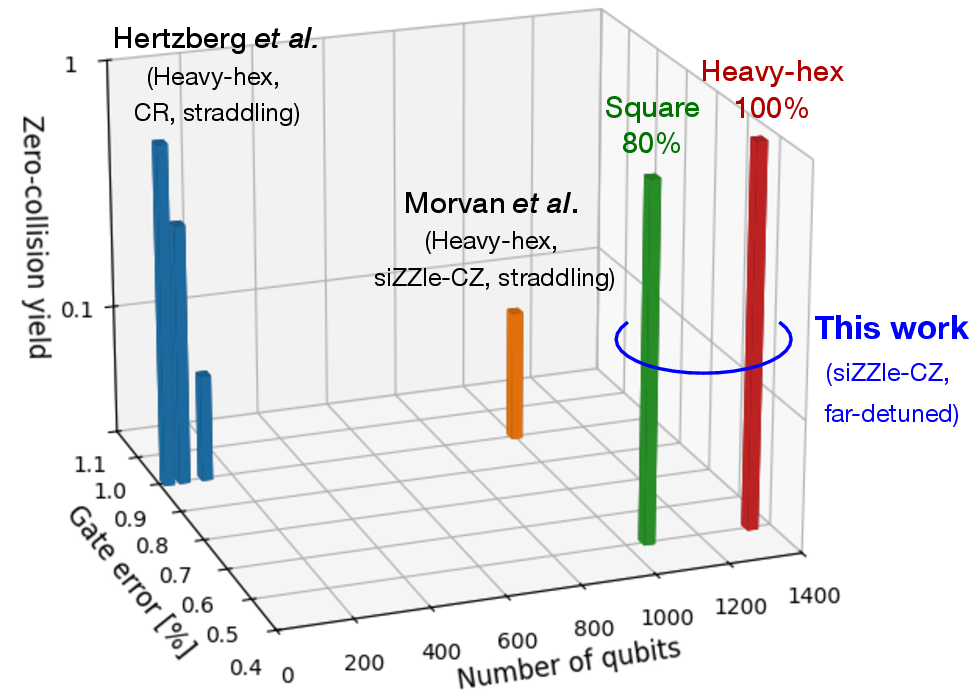}
\caption{
Position of our results relative to previous estimates of zero-collision yield for integrated qubit systems based on CR gates~\cite{hertzberg2021laser} and siZZle-CZ gates in the straddling regime~\cite{morvan2022optimizing}.
The present approach achieves higher zero-collision yields under conditions involving both larger system sizes and more stringent target gate errors.
}
\label{fig:compare}
\end{figure}

In this study, we apply the siZZle-CZ gate to fixed-frequency transmon qubits and systematically evaluate the gate fidelity and frequency collisions over a wide range of frequency detunings, including the far-detuned regime that has not been thoroughly explored in previous studies.
The results show that in the far-detuned regime, the allowable ranges of both the qubit and drive detunings are substantially enlarged, enabling the realization of CZ gates with fidelities exceeding $99.6\%$ over a broad parameter space.
This behavior indicates a high degree of robustness of the proposed approach against fabrication-induced dispersion in qubit frequencies.
Indeed, based on the conditions identified in this study, for chip designs with $>1000$ qubits, we demonstrate that zero-collision yields of $80\%$ and $100\%$ can be achieved for square and heavy-hexagonal lattices, respectively, by assuming a frequency dispersion of $0.25\%$.
Compared with the results obtained using architectures employing a CR gate as the 2Q gate \cite{hertzberg2021laser} and those obtained using approaches based on the siZZle-CZ gate in the straddling regime \cite{morvan2022optimizing}, our results demonstrate a distinct advantage from the perspective of scalability (Fig.~\ref{fig:compare}).
Moreover, previously, error-correction schemes have been proposed that can tolerate a certain fraction of missing nodes or edges arising from frequency collisions without severely affecting the effective code distance \cite{debroy2024luci, leroux2024snakes}.
This perspective suggests that even when the frequency dispersion surpasses 0.25\%, the present approach is expected to enable large-scale quantum error correction with a high effective yield, particularly when assessed in terms of low-collision yield rather than strictly zero-collision yield.
The frequency-collision criterion introduced in this paper is based on explicit evaluations of gate errors and thus provides a quantitative and physically motivated benchmark that can serve as a practical guideline for guaranteeing target fidelities already at the design stage.

The primary advantage of the far-detuned regime in enhancing the fabrication yield lies in its capacity to sustain substantial detunings, between neighboring qubits as well as between the drive frequency and other relevant transitions; thus, unwanted excitations are effectively suppressed.

Compared with the straddling regime, this regime generally requires larger drive amplitudes and broader pulse bandwidths.
However, the drive amplitude $200\,\mathrm{MHz}$ and bandwidth $1.5\,\mathrm{GHz}$ assumed in this study are comparable to those obtained for commercially available microwave control hardware \cite{PRXQuantum.6.020345}.
In addition, the present evaluation of frequency collisions is based on an idealized circuit model, whereas in experimental devices, additional factors---including (i) parasitic capacitive coupling between NNN and more distant qubits and (ii) microwave crosstalk during drive operations---may affect performance and yield.
Factor (i) represents a static imperfection inherent to fixed-frequency transmon architectures and increases the probability of frequency collisions independently of microwave driving. Factor (ii) contributes to dynamic collision processes induced by microwave control.
However, in the siZZle-CZ gate, the drive frequency can be selected to avoid frequency collisions, suggesting that the impact of microwave crosstalk is expected to be less severe than that in CR-gate-based schemes, wherein the drive frequency is fixed by the target qubit frequency.

Collectively, these results provide design guidelines for enhancing large-scale integration, particularly for highly connected lattice geometries such as square lattices.
The present approach, which combines frequency-allocation strategies with appropriate gate-selection schemes, enables direct feedback between fabrication yield considerations and the design of fault-tolerant quantum architectures.
Future investigations can be focused on the experimental implementation of the siZZle-CZ gate in the far-detuned regime, characterization under simultaneous multiqubit operations, and direct comparisons between experimentally measured collision rates and those predicted using the proposed framework.
Further improvements may be achieved by combining the proposed approach with pulse-shaping techniques, dynamical decoupling, and other noise-mitigation strategies as well as by extending the analysis to other superconducting platforms such as fluxonium qubits.
Through these efforts, the presented design strategy is expected to contribute to the realization of sustained subthreshold operation for fault-tolerant quantum computation.

\section{Methods}

\subsection{Design and evaluation strategy for various parameters}

\begin{figure*}[t]		
\centering
\includegraphics[width=18cm]{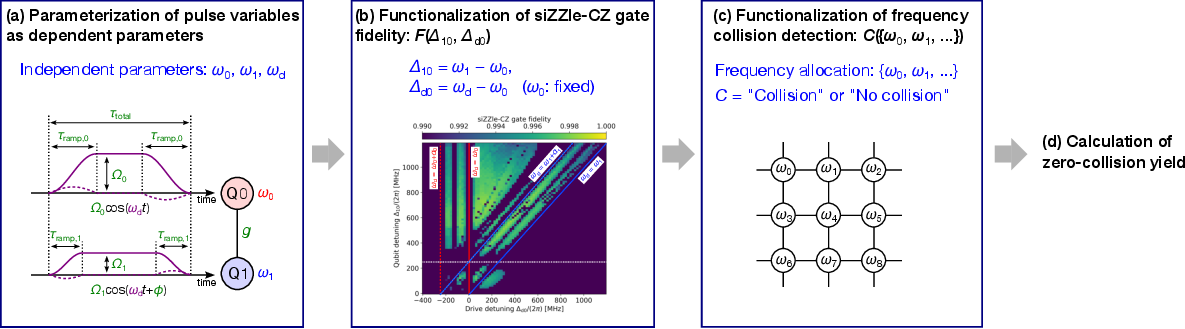}
\caption{
Hierarchical framework for the design of integrated qubit systems based on the siZZle-CZ gate and for the evaluation of the zero-collision yield.
(a) For each individual siZZle-CZ gate, optimal siZZle pulse waveforms are defined, such that multiple pulse parameters are treated as dependent variables.
(b) The siZZle-CZ gate fidelity is evaluated as a function of the qubit detuning $\varDelta_{10}$ and the drive detuning $\varDelta_{\mathrm{d0}}$, yielding a gate-fidelity function $F(\varDelta_{10}, \varDelta_{\mathrm{d0}})$.
(c) For a given frequency allocation across a lattice, the fidelity function $F(\varDelta_{10}, \varDelta_{\mathrm{d0}})$ is used to construct an evaluation function $C(\{\omega_0, \omega_1, \cdots\})$ that determines the presence of frequency collisions.
(d) The collision-evaluation function $C(\{\omega_0, \omega_1, \cdots\})$ is applied to various frequency allocations generated under stochastic frequency variations, from which the zero-collision yield is estimated.
}
\label{fig:sizzle_sekkei}
\end{figure*}

The objective of this study is to design an integrated superconducting qubit architecture that maximizes the zero-collision yield while targeting as high a siZZle-CZ gate fidelity as possible.
The primary design parameters to be determined are the resonance frequencies of individual qubits and the qubit--qubit coupling strengths.
However, the fidelity of the siZZle-CZ gate depends on the drive frequency as well as on the detailed shape of the siZZle pulse waveform, and the conditions, under which frequency collisions occur, likewise vary with the siZZle driving parameters.
In such a setting, wherein a large number of parameters are mutually interdependent, an efficient and systematic design methodology for integrated qubit chips is required.
Thus, we adopt the hierarchical evaluation framework illustrated in Fig.~\ref{fig:sizzle_sekkei}, in which the number of independent variables is progressively reduced at each layer, and the resulting evaluation functions are passed to the subsequent layer.

In the first stage, as shown in Fig.~\ref{fig:sizzle_sekkei}(a), we consider individual implementations of the siZZle-CZ gate and assume that the gate error originates solely from decoherence and nonadiabatic excitations.
For given qubit resonance frequencies $\omega_0$ and $\omega_1$, siZZle drive frequency $\omega_{\mathrm{d}}$, and qubit--qubit coupling strength $g$, the siZZle drive pulse is uniquely determined by imposing the condition that these error contributions are minimized.
Specifically, the siZZle pulse waveform is selected to maximize the pulse area under the constraint that the nonadiabatic error remains below $10^{-4}$ (see Sec.~\ref{sec:sizzle_params} for details).
This procedure allows the multiple waveform parameters characterizing the siZZle pulse to be treated as dependent variables.
In the second stage, illustrated in Fig.~\ref{fig:sizzle_sekkei}(b), we fix $\omega_0$ and $g$, and take the qubit detuning $\varDelta_{10}$ and the drive detuning $\varDelta_{\mathrm{d0}}$, which are defined as the deviations of $\omega_1$ and $\omega_{\mathrm{d}}$ from $\omega_0$, as independent variables.
For each pair $(\varDelta_{10}, \varDelta_{\mathrm{d0}})$, the siZZle-CZ gate fidelity is computed.
As a result, the gate fidelity can be expressed as a function $F(\varDelta_{10}, \varDelta_{\mathrm{d0}})$ of these two detuning parameters, as described in Sec.~\ref{sec:sizzle_calc}.
In the third stage, shown in Fig.~\ref{fig:sizzle_sekkei}(c), we consider a specific assignment of resonance frequencies to qubits on a lattice.
Using the fidelity function $F(\varDelta_{10}, \varDelta_{\mathrm{d0}})$, we evaluate the gate errors associated with each node and each edge of the lattice.
Based on these evaluations, we construct a function $C({\omega_0, \omega_1, \cdots})$ that outputs whether frequency collisions occur for the given frequency allocation (see Sec.\ref{sec:collision} for details).
Finally, in the last stage illustrated in Fig.~\ref{fig:sizzle_sekkei}(d), we evaluate the collision function $C({\omega_0, \omega_1, \cdots})$ for a large ensemble of frequency allocations generated by stochastic frequency variations.
From this ensemble, we estimate the probability of no frequency collision occurrence anywhere in the system, namely, the zero-collision yield.

\subsection{Operating principle of siZZle}\label{sec:sizzle}

\begin{figure}[t]		
\centering
\includegraphics[width=8cm]{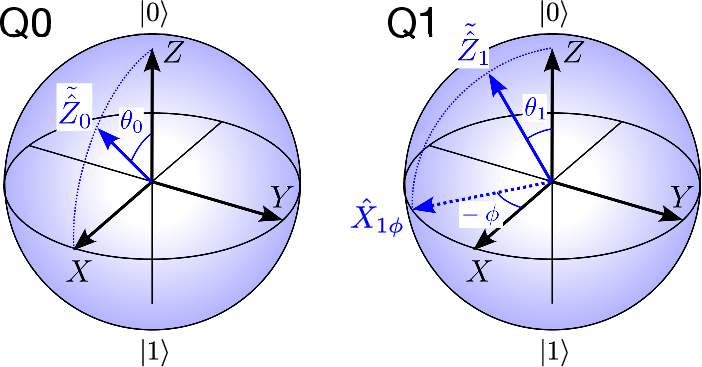}
\caption{
Geometric relations among the relevant operators on the Bloch spheres of Q0 and Q1 during siZZle pulse driving.
}
\label{fig:sizzle_bloch}
\end{figure}

SiZZle (Stark-induced ZZ by level excursions) is a method for inducing an effective ZZ interaction between two weakly coupled qubits by simultaneously applying off-resonant microwave drives at a common frequency $\omega_{\rm d}$ to both qubits.
While the basic operating principle of siZZle has been discussed in previous studies~\cite{wei2022hamiltonian, mitchell2021hardware}, herein, we revisit the mechanism from a complementary perspective, as siZZle plays a central role as the 2Q gate in this study.

For clarity, we first consider the case of two-level systems.
As shown in Fig.~\ref{fig:1}(a), the Hamiltonian of two coupled qubits under siZZle driving can be written as 
\begin{align}
\hat{H} &= -\omega_0\frac{\hat{Z}_0}{2}
-\omega_1\frac{\hat{Z}_1}{2}
+g(\hat{a}_0\hat{a}_1^\dag+\hat{a}_0^\dag\hat{a}_1)\no\\
&\quad + \varOmega_0\cos(\omega\sub{d}t)(\hat{a}_0+\hat{a}_0^\dag)
+\varOmega_1\cos(\omega\sub{d}t+\phi)(\hat{a}_1+\hat{a}_1^\dag),
\end{align}
where $i=0,1$ labels the qubits, $\hat{Z}_i$ denotes the Pauli-Z operator acting on Q-$i$, $\hat{a}_i:=\ket{0}_i\bra{1}_i$ is the annihilation operator for the corresponding two-level system, and we set $\hbar = 1$.
Applying the unitary transformation
\begin{align}
\hat{U}\sub{r}=\exp\left[\frac{-\ii\omega\sub{d}t(\hat{Z}_0+\hat{Z}_1)}{2}\right]
\end{align}
to move to the rotating frame, rapidly oscillating terms at frequency $2\omega_{\rm d}$ can be neglected under the rotating-wave approximation and the resulting effective Hamiltonian is given by
\begin{align}
\hat{H}\sub{eff} &= -\varDelta\sub{0d}\frac{\hat{Z}_0}{2}
+\varOmega_0\frac{\hat{X}_0}{2}
-\varDelta\sub{1d}\frac{\hat{Z}_1}{2}
+\varOmega_1\frac{\hat{X}_{1\phi}}{2}\no\\
&\quad 
+g(\hat{a}_0\hat{a}_1^\dag+\hat{a}_0^\dag\hat{a}_1),
\end{align}
where $\varDelta_{i{\rm d}}:=\omega_i-\omega\sub{d}$ ($i=0,1$). 
Here, $\hat{X}_i$ and $\hat{Y}_i$ denote the Pauli-X and Pauli-Y operators acting on Q-$i$, respectively, and $\hat{X}_{1\phi}:=\hat{X}_1\cos\phi-\hat{Y}_1\sin\phi$ represents a rotation of $\hat{X}_1$ by an angle $\phi$ in the equatorial plane.
We further define
\begin{align}
\tilde{\hat{Z}}_0 &:= \hat{Z}_0\cos\theta_0 + \hat{X}_0\sin\theta_0,\\
\tilde{\hat{Z}}_1 &:= \hat{Z}_1\cos\theta_1 + \hat{X}_{1\phi}\sin\theta_1,\\
\cos\theta_i &:= \frac{\varDelta_{i{\rm d}}}{\tilde{\omega}_i},\\
\sin\theta_i &:= \frac{-\varOmega_i}{\tilde{\omega}_i},\\
\tilde{\omega}_i &:= \sqrt{\varDelta_{i{\rm d}}^2+\varOmega_i^2},
\end{align}
which allows the effective Hamiltonian to be expressed as
\begin{align}
\hat{H}\sub{eff} &= -\tilde{\omega}_0\frac{\tilde{\hat{Z}}_0}{2}
- \tilde{\omega}_1\frac{\tilde{\hat{Z}}_1}{2}
+ \zeta\sub{siZZle}\frac{\tilde{\hat{Z}}_0\tilde{\hat{Z}}_1}{4}
+ \hat{C}\sub{\text{non-diag}},\\
\zeta\sub{siZZle} &:= 2g\cos\phi\sin\theta_0\sin\theta_1.
\end{align}
Here, $\hat{C}\sub{\text{non-diag}}$ contains only off-diagonal elements and has a norm of order $g$.
The geometric relations among the operators introduced above are illustrated on the Bloch spheres in Fig.~\ref{fig:sizzle_bloch}.
Provided that the ramp-up of the siZZle pulse is sufficiently slow, the operators $\tilde{\hat{Z}}_0$ and $\tilde{\hat{Z}}_1$ serve as adiabatically evolving quantization axes during the siZZle drive.
Applying a Schrieffer--Wolff transformation to $\hat{H}_{\rm eff}$, the off-diagonal terms can be suppressed to second order in $g$, while the $\tilde{\hat{Z}}_0\tilde{\hat{Z}}_1$ term remains at first order.
In this way, an effective ZZ interaction between the 2Qs is generated by the siZZle pulse.

In the regime $\varOmega_i \ll |\varDelta_{i{\rm d}}|$ $(i=0,1)$, the siZZle-induced ZZ interaction strength $\zeta\sub{siZZle}$ can be approximated as
\begin{align}
\zeta\sub{siZZle} \approx 2g\cos\phi
\frac{\varOmega_0}{\varDelta\sub{0d}}
\frac{\varOmega_1}{\varDelta\sub{1d}},
\end{align}
indicating that $\zeta_{\rm siZZle}$ increases proportionally with the drive amplitudes $\varOmega_i$.
This expression is consistent with Eq.~(S14) in the Supplemental Material of Ref.~\cite{wei2022hamiltonian}.
For larger $\varOmega_i$ or smaller detunings $|\varDelta_{i{\rm d}}|$, this approximation no longer holds, and $\zeta_{\rm siZZle}$ saturates before reaching $2g \cos\phi$.
In the numerical simulations presented in this paper, we also consider parameter regimes in which the condition $\varOmega_i \ll |\varDelta_{i{\rm d}}|$ is not necessarily satisfied.

In realistic devices, transmon qubits are weakly anharmonic multilevel systems.
To formulate a more accurate description of the siZZle-induced ZZ interaction between qubits, at least a three-level model must be considered.
The Hamiltonian of two coupled transmons under siZZle driving can be written as
\begin{align}
\hat{H} &= \sum_{i=0,1}
\left(
\omega_i\hat{a}_i^\dag\hat{a}_i + \frac{\alpha_i}{2}\hat{a}_i^{\dag 2}\hat{a}_i^2
\right)
+g(\hat{a}_0\hat{a}_1^\dag+\hat{a}_0^\dag\hat{a}_1)\no\\
&\quad + \varOmega_0\cos(\omega\sub{d}t)(\hat{a}_0+\hat{a}_0^\dag)
+\varOmega_1\cos(\omega\sub{d}t+\phi)(\hat{a}_1+\hat{a}_1^\dag),
\end{align}
where $\hat{a}_i$ denotes the annihilation operator for Q-$i$.
We choose the unitary transformation
\begin{align}
\hat{U}\sub{r} = \exp\left[
\ii\omega\sub{d}t(\hat{a}_0^\dag\hat{a}_0 + \hat{a}_1^\dag\hat{a}_1)
\right],
\end{align}
which yields the effective Hamiltonian in the rotating frame as
\begin{align}
\hat{H}\sub{eff} &=
\varDelta_{0{\rm d}}\hat{a}_0^\dag\hat{a}_0
+ \frac{\alpha_0}{2}\hat{a}_0^{\dag 2}\hat{a}_0^2
+\varOmega_0\frac{\hat{a}_0+\hat{a}_0^\dag}{2}
\no\\
&\quad
+\varDelta_{1{\rm d}}\hat{a}_1^\dag\hat{a}_1
+ \frac{\alpha_1}{2}\hat{a}_1^{\dag 2}\hat{a}_1^2
+\varOmega_1\frac{\ee^{\ii\phi}\hat{a}_1+\ee^{-\ii\phi}\hat{a}_1^\dag}{2}
\no\\
&\quad 
+g(\hat{a}_0\hat{a}_1^\dag+\hat{a}_0^\dag\hat{a}_1),
\end{align}
where $\varDelta_{i{\rm d}} := \omega_i - \omega_{\rm d}$.
Truncating this Hamiltonian to the three lowest levels of each transmon and expressing it in matrix form, we obtain
\begin{widetext}
\begin{align}
\hat{H}\sub{eff} &= 
\left(
\begin{bmatrix}
0 && \\
&\varDelta\sub{0d} &\\
&& 2\varDelta\sub{0d}+\alpha_0\\
\end{bmatrix}
+
\frac{\varOmega_0}{2}
\begin{bmatrix}
& 1 & \\
1 && \sqrt{2}\\
& \sqrt{2}& \\
\end{bmatrix}
\right)
\otimes \hat{I}
+ 
\hat{I}\otimes
\left(
\begin{bmatrix}
0 && \\
&\varDelta\sub{1d} &\\
&& 2\varDelta\sub{1d}+\alpha_1\\
\end{bmatrix}
+
\frac{\varOmega_1}{2}
\begin{bmatrix}
& \ee^{\ii\phi} & \\
\ee^{-\ii\phi} && \sqrt{2}\ee^{\ii\phi}\\
& \sqrt{2}\ee^{-\ii\phi}& \\
\end{bmatrix}
\right)\no\\
&\quad +
g\left(
\begin{bmatrix}
& 1 & \\
0&& \sqrt{2}\\
&0& \\
\end{bmatrix}
\otimes
\begin{bmatrix}
&0& \\
1 &&0\\
& \sqrt{2}& \\
\end{bmatrix}
+\Hc
\right).
\end{align}
Assuming that the siZZle drive amplitudes $\varOmega_i$ and the qubit--qubit coupling strength $g$ are sufficiently small compared with the detunings $\varDelta_{i{\rm d}}$ ($i=0,1$), we apply a Schrieffer--Wolff transformation and the resulting effective Hamiltonian is approximately given by
\begin{align}
& \hat{H}\sub{eff}' \approx
\begin{bmatrix}
-\frac{\varOmega_0^2}{2\varDelta\sub{0d}} & & \frac{\varOmega_0^2}{2\sqrt{2}}\left(-\frac{1}{\varDelta\sub{0d}}+\frac{1}{\varDelta\sub{0d}+\alpha_0}\right) \\
&\varDelta\sub{0d} + \frac{\varOmega_0^2}{2\varDelta\sub{0d}} - \frac{\varOmega_0^2}{\varDelta\sub{0d}+\alpha_0}  &\\
\frac{\varOmega_0^2}{2\sqrt{2}}\left(-\frac{1}{\varDelta\sub{0d}}+\frac{1}{\varDelta\sub{0d}+\alpha_0}\right) && 2\varDelta\sub{0d}+\alpha_0 + \frac{\varOmega_0^2}{\varDelta\sub{0d}+\alpha_0}\\
\end{bmatrix}
\otimes \hat{I}\no\\
&
+ 
\hat{I} \otimes 
\begin{bmatrix}
-\frac{\varOmega_1^2}{2\varDelta\sub{1d}} & & \frac{\varOmega_1^2}{2\sqrt{2}}\left(-\frac{1}{\varDelta\sub{1d}}+\frac{1}{\varDelta\sub{1d}+\alpha_1}\right)\ee^{\ii 2\phi} \\
&\varDelta\sub{1d} + \frac{\varOmega_1^2}{2\varDelta\sub{1d}} - \frac{\varOmega_1^2}{\varDelta\sub{1d}+\alpha_1}  &\\
\frac{\varOmega_1^2}{2\sqrt{2}}\left(-\frac{1}{\varDelta\sub{1d}}+\frac{1}{\varDelta\sub{1d}+\alpha_1}\right) \ee^{-\ii 2\phi} && 2\varDelta\sub{1d}+\alpha_1 + \frac{\varOmega_1^2}{\varDelta\sub{1d}+\alpha_1}\\
\end{bmatrix}
\no\\
&
+
\frac{g\varOmega_0\varOmega_1}{2}
\left(
\begin{bmatrix}
-\frac{1}{\varDelta\sub{0d}}& & \frac{\sqrt{2}}{\varDelta\sub{0d}+\alpha_0}-\frac{\sqrt{2}}{\varDelta\sub{0d}} \\
& -\frac{2}{\varDelta\sub{0d}+\alpha_0}+\frac{1}{\varDelta\sub{0d}}& \\
&& \frac{2}{\varDelta\sub{0d}+\alpha_0} \\
\end{bmatrix}
\otimes
\begin{bmatrix}
-\frac{1}{\varDelta\sub{1d}}\ee^{\ii\phi}& & \\
& \left(-\frac{2}{\varDelta\sub{1d}+\alpha_1}+\frac{1}{\varDelta\sub{1d}} \right)\ee^{\ii\phi}& \\
\left( \frac{\sqrt{2}}{\varDelta\sub{1d}+\alpha_1}-\frac{\sqrt{2}}{\varDelta\sub{1d}} \right)\ee^{-\ii\phi} && \frac{2}{\varDelta\sub{1d}+\alpha_1}\ee^{\ii\phi} \\
\end{bmatrix}
+\Hc
\right).
\end{align}
Further truncating this Hamiltonian to the computational subspace spanned by the two lowest levels of each transmon, we obtain
\begin{align}
\hat{H}\sub{eff}' &\approx
\begin{bmatrix}
-\frac{\varOmega_0^2}{2\varDelta\sub{0d}} &  \\
&\varDelta\sub{0d} + \frac{\varOmega_0^2}{2\varDelta\sub{0d}} - \frac{\varOmega_0^2}{\varDelta\sub{0d}+\alpha_0}  \\
\end{bmatrix}
\otimes \hat{I}
+ 
\hat{I} \otimes 
\begin{bmatrix}
-\frac{\varOmega_1^2}{2\varDelta\sub{1d}} & \\
&\varDelta\sub{1d} + \frac{\varOmega_1^2}{2\varDelta\sub{1d}} - \frac{\varOmega_1^2}{\varDelta\sub{1d}+\alpha_1} \\
\end{bmatrix}
\no\\
&
+
\frac{g\varOmega_0\varOmega_1\cos\phi}{2}
\begin{bmatrix}
\frac{1}{\varDelta\sub{0d}\varDelta\sub{1d}} &&& \\
& -\frac{1}{\varDelta\sub{0d}}\left(
-\frac{2}{\varDelta\sub{1d}+\alpha_1}+\frac{1}{\varDelta\sub{1d}}
 \right) && \\
 && -\frac{1}{\varDelta\sub{1d}}\left(
-\frac{2}{\varDelta\sub{0d}+\alpha_0}+\frac{1}{\varDelta\sub{0d}}
 \right) & \\
 &&&
 \left(
-\frac{2}{\varDelta\sub{0d}+\alpha_0}+\frac{1}{\varDelta\sub{0d}}
 \right)
 \left(
-\frac{2}{\varDelta\sub{1d}+\alpha_1}+\frac{1}{\varDelta\sub{1d}}
 \right)
 \\
\end{bmatrix}.
\end{align}
\end{widetext}
The strength of the siZZle-induced ZZ interaction is then obtained as
\begin{align}
\zeta\sub{siZZle} &= E_{00} + E_{11} - E_{01} - E_{10}\no\\
&= \frac{2g\varOmega_0\varOmega_1\alpha_0\alpha_1\cos\phi}{\varDelta\sub{0d}(\varDelta\sub{0d}+\alpha_0)\varDelta\sub{1d}(\varDelta\sub{1d}+\alpha_1)},
\end{align}
which coincides with Eq.~(6) of Ref.~\cite{wei2022hamiltonian}.
As in the two-level analysis, this expression is valid only in the regime where $\varOmega_i$ is sufficiently small compared with both $|\varDelta_{i{\rm d}}|$ and $|\varDelta_{i{\rm d}} + \alpha_i|$.
In general, the siZZle-induced ZZ interaction saturates before reaching the value $2g \cos\phi$.

\subsection{Numerical simulation of quantum dynamics}\label{sec:qutip}

In this study, numerical simulations of the quantum dynamics were performed using QuTiP \cite{lambert2024qutip,qutip}. 
The Hamiltonian assumed in the simulations is given by (we set $\hbar = 1$)
\begin{align}
    \hat{H}&:=\sum_{i=0,1}\hat{H}_i+\sum_{i=0,1}\hat{H}_{{\rm d},i}(t)+\hat{H}_{\rm int},\\
    \hat{H}_i &:=\omega_i \hat{a}_i\hat{a}_i^\dag + \frac{\alpha_i}{2}\hat{a}_i^{\dag 2}\hat{a}_i^2, \\
    \hat{H}_{{\rm d},i}(t) &:= \varOmega_i(t)\cos\left[\omega_{{\rm d},i}t+\phi_i(t)\right](\hat{a}_i+\hat{a}_i^\dag),\\
    \hat{H}_{\rm int} &:= g(\hat{a}_0\hat{a}_1^\dag + \Hc),
\end{align}
Here, $i=0,1$ labels the qubits, and $\hat{a}_i$ denotes the annihilation operator of Q-$i$.
$\hat{H}_i$ is the free Hamiltonian of qubit $i$, where $\omega_i$ and $\alpha_i$ are the g-e transition frequency and anharmonicity, respectively.
Unless otherwise specified, we fix the parameters of Q0 to
$\omega_0 = 2\pi \times 5000~{\rm MHz}$ and 
$\alpha_0 = -0.05\,\omega_0 = -2\pi \times 250~{\rm MHz}$,
corresponding approximately to $E_J/E_C \approx 50$.
For Q1, $\omega_1$ and $\alpha_1 = -0.05\,\omega_1$ are treated as sweep parameters.
$\hat{H}_{{\rm d},i}(t)$ describes the microwave drive applied to Q-$i$, where $\varOmega_i(t)$ and $\phi_i(t)$ are the time-dependent drive amplitude and phase, respectively, and $\omega_{{\rm d},i}$ is the drive frequency.
$\hat{H}_{\rm int}$ represents the transverse coupling between the 2Qs with coupling strength $g$.

For numerical efficiency, we transform the Hamiltonian into a rotating frame with frame frequency
$\omega_{\rm f} = 2\pi \times 5600~{\rm MHz}$
applied to both qubits, thereby reducing fast time dependence. 
The transformed Hamiltonian reads
\begin{align}
    \hat{H}\sub{f}&=\sum_{i=0,1}\hat{H}_{{\rm f},i}+\sum_{i=0,1}\hat{H}_{{\rm df},i}(t)+\hat{H}_{\rm int},\\
    \hat{H}_{{\rm f},i} &=(\omega_i-\omega\sub{f}) \hat{a}_i\hat{a}_i^\dag + \frac{\alpha_i}{2}\hat{a}_i^{\dag 2}\hat{a}_i^2, \\
    \hat{H}_{{\rm df},i}(t) &= 
    {\rm Re}\left[
    \varOmega_i(t)\ee^{\ii\phi_i(t)}\ee^{\ii (\omega\sub{d}-\omega\sub{f})t}
    \right]
    \frac{\hat{a}_i+\hat{a}_i^\dag}{2}\no\\
    & \quad 
    - {\rm Im}\left[
    \varOmega_i(t)\ee^{\ii\phi_i(t)}\ee^{\ii (\omega\sub{d}-\omega\sub{f})t}
    \right]
    \frac{\hat{a}_i-\hat{a}_i^\dag}{2\ii},\\
    \hat{H}_{\rm int} &= g(\hat{a}_0\hat{a}_1^\dag + \Hc).
\end{align}
Finally, in the numerical simulations each transmon was truncated to the lowest four Fock states, and the discretized Hamiltonian was constructed in this finite-dimensional Hilbert space.

To simulate time evolution including relaxation processes, we solved the following Lindblad master equation using the $\tt mesolve$ function in QuTiP:
\begin{align}
\dot{\hat{\rho}}(t) &= -\frac{\ii}{\hbar}[\hat{H}(t), \hat{\rho}(t)]
+\mathcal{L}(\hat{\rho}),\\
\mathcal{L}(\hat{\rho}) &:=
\frac{1}{2}\sum_{j}
\left[
2\hat{L}_j\hat{\rho}(t)\hat{L}_j^\dag
- \hat{\rho}(t)\hat{L}_j^\dag\hat{L}_j
- \hat{L}_j^\dag\hat{L}_j\hat{\rho}(t)
\right].
\end{align}
The time step was set to 0.1\,ns. 
This value satisfies the condition that the time step be sufficiently smaller than the characteristic time scale of the Hamiltonian $\hat{H}_{\rm f}(t)$, whose fastest oscillation period is approximately $1/(500\,{\rm MHz}) \approx 2\,{\rm ns}$.
For simulations including relaxation, we assumed energy relaxation time $T_1 = 600\,\mu{\rm s}$ and pure dephasing time $T_2^* = 600\,\mu{\rm s}$ for both qubits. 
The corresponding Lindblad operators are
\begin{align}
\hat{L}_1 &= \sqrt{\gamma_1}(\hat{a}_0+\hat{a}_1),\\
\hat{L}_2 &= \sqrt{2\gamma\sub{ph}}(\hat{a}_0^\dag\hat{a}_0 + \hat{a}_1^\dag\hat{a}_1),\\
\gamma_1 &= \frac{1}{T_1},\\
\gamma\sub{ph} &= \frac{1}{T_2^*} - \frac{1}{T_1}.
\end{align}

Expectation values were computed using the QuTiP $\tt expect$ function.
Since the Hamiltonian is defined in the rotating frame with frequency $\omega_{\rm f}$, the expectation values of observables other than $\hat{Z}$ (e.g., $\hat{X}$ and $\hat{Y}$) generally differ from those defined in the qubit's laboratory-frequency frame.
To obtain expectation values in the qubit-frequency frame, we numerically calibrated a half-$\pi$ pulse and combined it with $\hat{Z}$ expectation-value measurements. 
In this way, measurements of $\hat{X}$ and $\hat{Y}$ in the qubit frame were effectively implemented.
The half-$\pi$ pulse used in the simulations is a cosine-shaped pulse with DRAG correction \cite{PhysRevLett.103.110501}, characterized by the complex envelope
\begin{align}
\tilde{\varOmega}_{{\rm hpi},i}(t) &= 
\varOmega_i
\left[
\frac{1 - \cos(2\pi t/\tau_{\rm hpi})}{2} 
+ \ii r_{{\rm iq},i} \frac{\sin(2\pi t/\tau_{\rm hpi})}{2}
\right],\no\\
&\hspace{4cm}(0 \leq t < \tau_{\rm hpi})
\end{align}
where $i=0,1$ labels the qubit, and the pulse duration was set to $\tau_{\rm hpi}=25\,{\rm ns}$.
The drive amplitude $\varOmega_i$ and the DRAG in-phase to quadrature (IQ) ratio $r_{{\rm iq},i}$ were numerically calibrated to realize an accurate half-$\pi$ rotation.

\subsection{Precalculation of the DRAG IQ ratio for siZZle pulses}\label{sec:rcft_drag}

For relatively long drive pulses such as those used in siZZle-CZ gates, we employed a raised-cosine flat-top (RCFT) pulse shape consisting of cosine-shaped ramp sections and a flat-top region. 
To suppress nonadiabatic transitions under off-resonant driving, we incorporated a DRAG waveform in the ramp sections of the RCFT pulse, where a sine-shaped quadrature (Q) component is added to the cosine-shaped in-phase (I) component. 
Denoting the ratio between the I and Q components (IQ ratio) by $r_{\rm iq}$, the value of $r_{\rm iq}$ that minimizes nonadiabatic transitions depends on the pulse-shape parameters and the drive frequency. 
In this section, we describe the procedure for determining the optimal $r_{\rm iq}$ for each parameter set using numerical simulations performed with QuTiP.

\begin{figure}[t]		
\centering
\includegraphics[width=8.6cm]{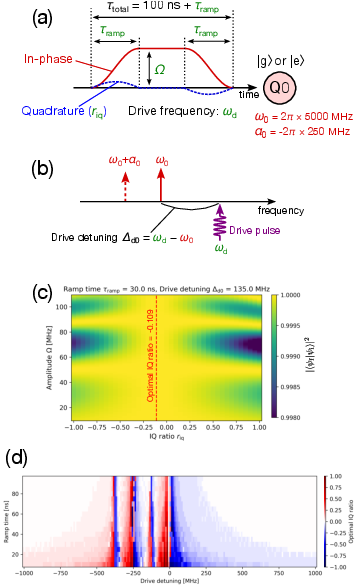}
\caption{
(a) Numerical simulation setup for determining the DRAG IQ ratio $r_{\rm iq}$ that suppresses nonadiabatic transitions in an RCFT pulse.
(b) Frequency configuration used in the simulations, showing the relationship between the g-e transition frequency $\omega_0$ of transmon Q0 and the drive frequency $\omega_{\rm d}$.
(c) Calculated projection probability $|\bracketi{\psi\sub{f}}{\psi\sub{i}}|^2$ as a function of $\varOmega$ and $r_{\rm iq}$ for $\ket{\psi\sub{i}}=\ket{\rm g}$, $\tau_{\rm ramp}=30\,{\rm ns}$, and $\varDelta_{\rm d0}=135\,{\rm MHz}$. The optimal value minimizing nonadiabatic transitions under this condition is $r_{\rm iq}=-0.109$.
(d) Distribution of the optimal $r_{\rm iq}$ obtained by averaging the independently optimized values for the initial states $\ket{\rm g}$ and $\ket{\rm e}$.
}
\label{fig:rcft_drag}
\end{figure}

Figure~\ref{fig:rcft_drag}(a) illustrates the simulation setup. 
The complex envelope of the RCFT pulse, $\tilde{\varOmega}_{\rm rcft}(t)$, is given by
\begin{widetext}
\begin{align}
\tilde{\varOmega}_{{\rm rcft}}(t) = 
\left\{
\begin{array}{ll}
\displaystyle
\varOmega 
\left[
\frac{1 - \cos(\pi t/\tau_{{\rm ramp}})}{2} + \ii r_{{\rm iq}} \frac{\sin(\pi t/\tau_{{\rm ramp}})}{2}
\right]
\quad & (0\leq t<\tau_{{\rm ramp}})\\
\displaystyle
\varOmega
& (\tau_{{\rm ramp}}\leq t<\tau_{\rm total}-\tau_{{\rm ramp}})\\
\displaystyle
\varOmega
\left\{
\frac{1 - \cos[\pi (\tau_{\rm total} - t)/\tau_{{\rm ramp}}]}{2} 
+ \ii r_{{\rm iq}} \frac{\sin[\pi (\tau_{\rm total} - t)/\tau_{{\rm ramp}}]}{2}
\right\}
\quad & (\tau_{\rm total}-\tau_{{\rm ramp}}\leq t<\tau_{\rm total})\\
\end{array}
\right.
\end{align}
\end{widetext}
where $\varOmega$, $\tau_{\rm ramp}$, and $\tau_{\rm total}$ denote the pulse amplitude, ramp duration, and total pulse duration, respectively. 
For a fixed $\varOmega$, we choose $\tau_{\rm total} = 100\,{\rm ns} + \tau_{\rm ramp}$ such that the pulse area of the real component remains constant irrespective of $\tau_{\rm ramp}$.
Figure~\ref{fig:rcft_drag}(b) shows the frequency configuration used in the simulations for a single transmon qubit (Q0). 
The g-e transition frequency and anharmonicity of Q0 are fixed to $\omega_0/(2\pi)=5000\,{\rm MHz}$ and $\alpha_0/(2\pi)=-250\,{\rm MHz}$, respectively. 
For a drive frequency $\omega_{\rm d}$, the drive detuning is defined as $\varDelta_{\rm d0} := \omega_{\rm d} - \omega_0$. 
In the numerical simulations, the initial state of Q0, $\ket{\psi\sub{i}}$, is prepared in either $\ket{\rm g}$ or $\ket{\rm e}$. 
We then sweep $r_{\rm iq}$, $\varOmega$, $\tau_{\rm ramp}$, and $\varDelta_{\rm d0}$ as control parameters. 
For each parameter set, the final state $\ket{\psi\sub{f}}$ after the pulse driving is computed, and the projection probability $|\bracketi{\psi\sub{f}}{\psi\sub{i}}|^2$ is evaluated.

As a representative example, Fig.~\ref{fig:rcft_drag}(c) shows the simulation result of $|\bracketi{\psi\sub{f}}{\psi\sub{i}}|^2$ as a function of $\varOmega$ and $r_{\rm iq}$ for $\ket{\psi\sub{i}}=\ket{\rm g}$, $\tau_{\rm ramp}=30\,{\rm ns}$, and $\varDelta_{\rm d0}=135\,{\rm MHz}$. 
For all values of $\varOmega$, the projection probability is maximized at the same value of $r_{\rm iq}$ (in this case, $r_{\rm iq}=-0.109$). 
This value therefore corresponds to the optimal $r_{\rm iq}$ that minimizes nonadiabatic transitions under the given conditions.

We determined the optimal $r_{\rm iq}$ for various combinations of $\tau_{\rm ramp}$ and $\varDelta_{\rm d0}$ for both initial states $\ket{\rm g}$ and $\ket{\rm e}$. 
Since a general qubit state is an arbitrary superposition of $\ket{\rm g}$ and $\ket{\rm e}$, we define the optimal $r_{\rm iq}$ used in the experiment as the average of the optimal values obtained for the two initial states. 
Figure~\ref{fig:rcft_drag}(d) presents the resulting distribution of the averaged optimal $r_{\rm iq}$. 
The sign of the optimal $r_{\rm iq}$ reverses as the drive detuning $\varDelta_{\rm d0}$ crosses the g-e, g-f, e-f, and e-h transition frequencies. 
Moreover, shorter ramp times $\tau_{\rm ramp}$ require larger magnitudes of $r_{\rm iq}$ to suppress nonadiabatic transitions effectively.

\subsection{Characterization of nonadiabatic transition probability via the adiabatic coefficient}

We numerically investigate how the nonadiabatic transition probability induced in a transmon under off-resonant pulse driving, such as the siZZle pulse, can be characterized by pulse parameters.  
According to the quantum adiabatic theorem, even if a Hamiltonian varies in time, a system initially prepared in an eigenstate of the Hamiltonian remains in the corresponding instantaneous eigenstate provided that the temporal variation is sufficiently slow.  
For an off-resonant pulse drive applied to a transmon qubit, if the temporal variation of the pulse amplitude is sufficiently slow, then a system prepared initially in $\ket{\rm g}$ or $\ket{\rm e}$ remains in $\ket{\rm g}$ or $\ket{\rm e}$, respectively, after the pulse; namely, nonadiabatic transitions are suppressed.  
For a general superposition state $\alpha\ket{\rm g}+\beta\ket{\rm e}$, the state after the pulse becomes $\alpha\ket{\rm g}+\ee^{\ii\phi}\beta\ket{\rm e}$: the population weights remain unchanged, while only the relative phase $\phi$ is modified. 
This phase shift can be compensated by a virtual-$Z$ gate.

The adiabaticity of a time-dependent Hamiltonian, namely, the degree of slowness of its temporal variation, is generally characterized by the adiabatic coefficient.  
Let $E_m(t)$ and $\ket{m(t)}$ denote the $m$-th instantaneous eigenvalue and eigenstate of the time-dependent Hamiltonian $\hat{H}(t)$ at time $t$, respectively.  
For a transmon qubit, the total adiabatic coefficient $A_{\rm total}$ is defined as \cite{Sakurai_Napolitano_2020}
\begin{align}
A_{\rm total} &= \sum_{(m,n)} A_{mn}, \\
A_{mn} &= \frac{\left|\bracketii{m(t)}{\dot{\hat{H}}(t)}{n(t)}\right|}{\left|E_m(t)-E_n(t)\right|^2},
\end{align}
where the summation in the first equation is taken over $(m,n)=({\rm g},{\rm e}), ({\rm g},{\rm f}), ({\rm e},{\rm f}), ({\rm e},{\rm h})$.  
A smaller value of $A_{\rm total}$ compared with unity indicates a higher degree of adiabaticity, i.e., a lower nonadiabatic transition probability.

\begin{figure}[t]		
\centering
\includegraphics[width=8.6cm]{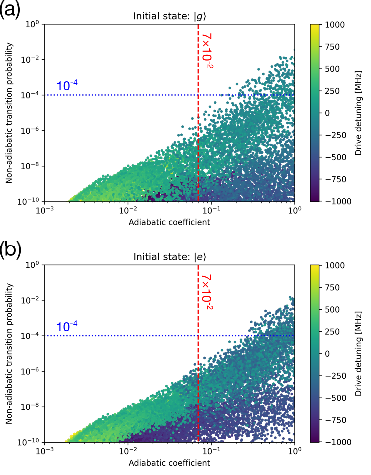}
\caption{
Numerical simulations of the relationship between the adiabatic coefficient calculated from each pulse waveform and the corresponding nonadiabatic transition probability $1-|\bracketi{\psi\sub{f}}{\psi\sub{i}}|^2$.
(a) For the initial state $\ket{\rm g}$, and (b) for the initial state $\ket{\rm e}$.
The red dashed line and the blue dotted line indicate the conditions $A_{\rm total}=7\times10^{-2}$ and $1-|\bracketi{\psi\sub{f}}{\psi\sub{i}}|^2=10^{-4}$, respectively.
The color of each data point represents the detuning $\varDelta_{\rm d0}$ between the drive frequency and the g-e transition frequency of transmon Q0.
In the vicinity of $\varDelta_{\rm d0}=0,{\rm MHz}$ and $250,{\rm MHz}$, where the drive frequency approaches the g-e and e-f transition frequencies, respectively, the nonadiabatic transition probability increases significantly. Correspondingly, the adiabatic coefficient also takes larger values in these regions.
}
\label{fig:adiab_coeff}
\end{figure}

The quantitative relationship between $A_{\rm total}$ and the nonadiabatic transition probability can be examined by analyzing the QuTiP simulations presented in Sec.~\ref{sec:rcft_drag}.  
In those simulations, we computed the adiabatic coefficient for each pulse waveform and plotted it against the corresponding nonadiabatic transition probability $1-|\bracketi{\psi\sub{f}}{\psi\sub{i}}|^2$, as shown in Fig.~\ref{fig:adiab_coeff}.  
These results, obtained within the numerical framework considered in this study, reveal that when the adiabatic coefficient calculated from the pulse parameters satisfies $A_{\rm total} \le 7 \times 10^{-2}$, the nonadiabatic transition probability drops below $10^{-4}$, independent of the initial state.

\subsection{Selection of pulse parameters in numerical simulations of the siZZle-CZ gate}\label{sec:sizzle_params}

In the numerical simulations of the siZZle-CZ gate presented in Sec.~\ref{sec:sizzle_calc}, we explain how the pulse waveform parameters, namely the ramp time $\tau_{{\rm ramp},i}$ and the drive amplitude $\varOmega_i$, are determined for a given drive detuning $\varDelta_{{\rm d}i}$ ($i=0,1$).
First, the optimal DRAG IQ ratio $r_{{\rm iq},i}$ is a function of $\varDelta_{{\rm d}i}$ and $\tau_{{\rm ramp},i}$. Therefore, when $\varDelta_{{\rm d}i}$ is fixed, $r_{{\rm iq},i}$ is uniquely determined once $\tau_{{\rm ramp},i}$ is specified.
Next, the adiabatic coefficient of transmon $i$, denoted $A_{{\rm total},i}$, is in general a function of $\varDelta_{{\rm d}i}$, $\tau_{{\rm ramp},i}$, $r_{{\rm iq},i}$, and $\varOmega_i$. However, when $\varDelta_{{\rm d}i}$ is fixed and $r_{{\rm iq},i}$ depends only on $\tau_{{\rm ramp},i}$, the adiabatic coefficient $A_{{\rm total},i}$ can be regarded as a function of $\tau_{{\rm ramp},i}$ and $\varOmega_i$.

\begin{figure}[t]		
\centering
\includegraphics[width=8cm]{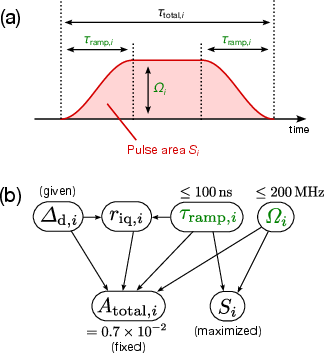}
\caption{
(a) Total pulse duration $\tau_{{\rm total},i}$, ramp time $\tau_{{\rm ramp},i}$, drive amplitude $\varOmega_i$, and pulse area $S_i$ for the siZZle pulse waveform.
(b) Parameter dependencies.
The four quantities listed above constitute the pulse waveform parameters. Among them, the two parameters highlighted in green, $\tau_{{\rm ramp},i}$ and $\varOmega_i$, are the variables to be optimized.
}
\label{fig:params}
\end{figure}

To maximize the fidelity of the siZZle-CZ gate, the nonadiabatic transitions induced by the off-resonant siZZle drive should be suppressed while increasing the $ZZ$ interaction strength through a large drive amplitude; this approach reduces the total gate time.
To satisfy the former requirement, we impose the constraint
$A_{{\rm total},i} = 0.7 \times 10^{-2}$,
which guarantees that the nonadiabatic transition probability remains below $10^{-4}$. Under this constraint, $\tau_{{\rm ramp},i}$ can be regarded as a function of $\varOmega_i$.
To satisfy the latter requirement, we consider maximizing the pulse area $S_i$, defined as the time integral of the in-phase component of the pulse waveform, as illustrated in Fig.~\ref{fig:params}(a). The pulse area is expressed as
\begin{align}
S_i = (\tau_{{\rm total},i} - \tau_{{\rm ramp},i}) \varOmega_i.
\end{align}
Accordingly, the optimal values of $\tau_{{\rm ramp},i}$ and $\varOmega_i$ are determined by maximizing $S_i$ under the imposed constraints.
In the numerical calculations presented in Sec.\ref{sec:sizzle_calc}, we adopt realistic upper bounds of $100\,{\rm ns}$ for $\tau_{{\rm ramp},i}$ and $200\,{\rm MHz}$ for $\varOmega_i$. The pulse parameters for each $\varDelta_{{\rm d}i}$ are therefore obtained by solving the following optimization problem:
\begin{align}
{\bf Maximize} & \quad S_i(\tau_{{\rm ramp},i}, \varOmega_i)\\
{\bf Subject\ to} & \quad A_{{\rm total},i} = 0.7\times 10^{-2}\\
& \quad \tau_{{\rm ramp},i}\leq 100\,{\rm ns}\\
& \quad \varOmega_i\leq 200\,{\rm MHz}
\end{align}
The resulting parameter dependencies are shown in Fig.~\ref{fig:params}(b).

\subsection{Calibration procedure for the siZZle-CZ gate}\label{sec:sizzle_calib}


We describe the procedure used to calibrate the siZZle-CZ gate in numerical simulations performed with QuTiP, including the determination of the total pulse duration $\tau_{\rm total}$ and the calibration of the virtual-$Z$ gates.
We first determine $\tau_{\rm total}$ using the following quantum circuit:
\begin{align}
\begin{quantikz}[thin lines]
  \lstick{Q0: $\ket{\rm g}$ or $\ket{\rm e}$} & \gate[wires=2][1.5cm]{{\rm siZZle}(\tau\sub{total})} & \qw  \\
  \lstick{Q1: $\ket{+}$} & \qw & \meter{\hat{X}_1\ {\rm or}\ \hat{Y}_1}
\end{quantikz}
\end{align}
Q0 is prepared in either $\ket{\rm g}$ or $\ket{\rm e}$, while Q1 is prepared in $\ket{+} := (\ket{\rm g} + \ket{\rm e})/\sqrt{2}$. After applying a siZZle pulse of duration $\tau_{\rm total}$, which induces a $ZZ$ interaction, we measure the expectation value of $\hat{X}_1$ or $\hat{Y}1$ on Q1.
For each initial state of Q0, the rotation angle of Q1 induced by the siZZle-induced $ZZ$ interaction, denoted $\phi_{{\rm siZZle},s}$ ($s = {\rm g}, {\rm e}$), is obtained from the measured expectation values $\langle \hat{X}_1 \rangle_s$ and $\langle \hat{Y}1 \rangle_s$ as
\begin{align}
\phi_{{\rm siZZle},s}=\arctan(\bracket{\hat{Y}_1}_s / \bracket{\hat{X}_1}_s).
\end{align}
The calibrated pulse duration $\tau\sub{total}\sur{cal}$ is chosen such that the phase difference
$\Delta \phi_{\rm siZZle}
:= \phi_{{\rm siZZle,g}} - \phi_{{\rm siZZle,e}}$
is equal to $\pi$.


Because the siZZle pulse is off-resonant for both Q0 and Q1, AC Stark shifts induce additional single-qubit $Z$ rotations on each transmon. These excess $Z$ rotations can be compensated by appropriate virtual-$Z$ gates, thereby realizing an overall CZ gate.
We initialize the virtual-$Z$ rotation angles as $\phi_{{\rm vz},i} = 0$ for $i=0,1$. From the circuit below, we determine the excess $Z$ rotation of Q1 under the siZZle drive by measuring $\langle \hat{X}_1 \rangle$ and $\langle \hat{Y}_1 \rangle$. 
The induced phase is obtained as $\arctan(\bracket{\hat{Y}_1}/\bracket{\hat{X}_1})$, and thus the correction angle for Q1 is $\phi_{{\rm vz},1}=-\arctan(\bracket{\hat{Y}_1}/\bracket{\hat{X}_1})$.
\begin{align}
\begin{quantikz}[thin lines]
  \lstick{Q0: $\ket{\rm g}$} & \gate[wires=2][1.5cm]{{\rm siZZle}(\tau\sub{total}\sur{cal})} & \gate{R_Z(\phi_{\rm vz,0})} & \qw  \\
  \lstick{Q1: $\ket{+}$} & \qw & \gate{R_Z(\phi_{\rm vz,1})} & \meter{\hat{X}_1\ {\rm or}\ \hat{Y}_1}
\end{quantikz}
\end{align}
Similarly, from the following circuit we determine the excess $Z$ rotation of Q0 under the siZZle drive by measuring $\langle \hat{X}_0 \rangle$ and $\langle \hat{Y}_0 \rangle$, yielding $\phi_{{\rm vz},0}=-\arctan(\bracket{\hat{Y}_0}/\bracket{\hat{X}_0})$.
\begin{align}
\begin{quantikz}[thin lines]
  \lstick{Q0: $\ket{+}$} & \gate[wires=2][1.5cm]{{\rm siZZle}(\tau\sub{total}\sur{cal})} & \gate{R_Z(\phi_{\rm vz,0})} & \meter{\hat{X}_0\ {\rm or}\ \hat{Y}_0}  \\
  \lstick{Q1: $\ket{\rm g}$} & \qw & \gate{R_Z(\phi_{\rm vz,1})} &
\end{quantikz}
\end{align}
With these procedures, the calibration of the siZZle-CZ gate is completed.

\subsection{Gate fidelity evaluation via quantum process tomography}\label{sec:qpt}

The fidelity of the siZZle-CZ gate in the QuTiP-based numerical simulation was evaluated by performing quantum process tomography (QPT) within the numerical framework.  
First, the Pauli transfer matrix (PTM) was obtained using the following quantum circuit:
\begin{align}
\begin{quantikz}[thin lines]
  \lstick{Q0: $\hat{P}_0$} & \gate[wires=2][1.5cm]{\text{siZZle-CZ}} & \qw & \meter{\hat{X}_0,\ \hat{Y}_0,\ {\rm or}\ \hat{Z}_0} \\
  \lstick{Q1: $\hat{P}_1$} & \qw & \qw & \meter{\hat{X}_1,\ \hat{Y}_1,\ {\rm or}\ \hat{Z}_1}
\end{quantikz}
\end{align}
Here, $\hat{P} := \hat{P}_0 \otimes \hat{P}_1$ denotes a 2Q Pauli operator.  
Although such operators do not correspond to physically allowed quantum states, they can be directly specified as initial states in numerical simulations using QuTiP.
The matrix element $M_{ij}$ of the PTM $M$ $(i,j \in \{II, \ldots, ZZ\})$ is obtained by preparing a 2Q Pauli operator $\hat{P}_i$ as the initial state, applying the siZZle-CZ gate under evaluation, $\mathcal{E}_{\rm CZ}$, and then measuring the expectation value of another 2Q Pauli operator $\hat{P}_j$ on the output state:
\begin{align}
M_{ij} = \frac{1}{2^n}\tr\left[\mathcal{E}\sub{CZ}(\hat{P}_i)\hat{P}_j\right],
\end{align}
where $n$ is the number of qubits ($n=2$ in this study).
Multiplying this matrix by the inverse of the PTM corresponding to the ideal CZ gate $M_{\rm ideal}$ yields the PTM $M_{\rm error}$, which contains only the error component of the evaluated siZZle-CZ gate:
\begin{align}
M\sub{error}=M M\sub{ideal}^{-1}.
\end{align}

Next, we compute the $\chi$ matrix corresponding to $M_{\rm error}$.  
The $\chi$ matrix is defined through the expansion of a quantum process $\mathcal{E}$ as
\begin{align}
 \mathcal{E}(\hat{\rho}) = 
 \sum_{mn}\chi_{mn}\hat{P}_m\hat{\rho}\hat{P}_n,
\end{align}
where $\chi_{mn}$ are the matrix elements of the $\chi$ matrix.
The $\chi$ matrix and the PTM are related by
\begin{align}
M_{ij}&= 
2^n
\sum_{mn}\chi_{mn}
A_{ij,mn},
\end{align}
where
\begin{align}
 A_{ij,mn} :=
\frac{
 {\rm tr}\left(
    \hat{P}_m\hat{P}_i\hat{P}_n\hat{P}_j
 \right)
 }
 {2^{2n}}
\end{align}
is a Hermitian and unitary matrix.
By inverting this relation, we obtain the $\chi$ matrix corresponding to $M_{\rm error}$, denoted as $\chi_{\rm error}$.  
The $(II,II)$ element of $\chi_{\rm error}$ is then taken as the fidelity of the evaluated siZZle-CZ gate.

\subsection{Determination of the qubit--qubit coupling strength $g$}\label{sec:g}

We describe the procedure used to determine the qubit--qubit coupling strength $g$ employed in the present numerical simulations.  
A larger coupling strength enhances the ZZ interaction amplified by the siZZle drive, thereby shortening the CZ gate duration and improving the gate fidelity under finite coherence times.  
In a multiqubit architecture, increasing $g$ also strengthens unwanted interactions with neighboring spectator qubits, broadening the frequency-collision conditions. Consequently, a larger $g$ does not necessarily improve the overall chip performance.
To quantify this trade-off, we numerically evaluated the idle-time evolution of two qubits coupled with strength $g$ using QuTiP, and investigated the dependence of spectator-induced errors on $g$.

\begin{figure}[t]		
\centering
\includegraphics[width=9cm]{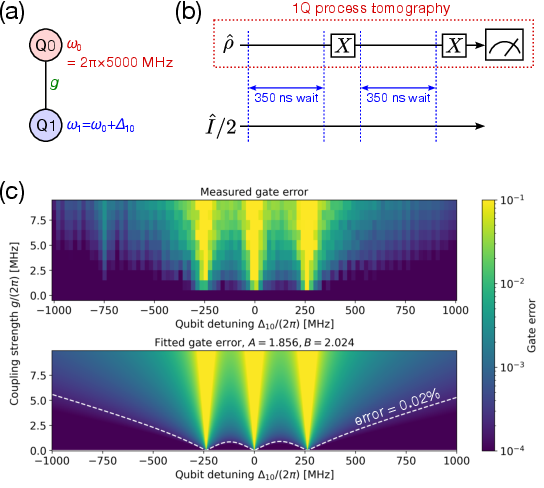}
\caption{
(a) Numerical setup used to determine an appropriate value of the qubit--qubit coupling strength $g$.
The coupling strength $g$ and the qubit detuning $\varDelta_{10}$ were treated as sweep parameters.
(b) Quantum circuit representing the time evolution evaluated in the numerical simulation.
(c) Upper panel: numerically computed idling error $E_{\rm idle}$ as functions of $\varDelta_{10}$ and $g$.
Lower panel: corresponding fitting results.
The regions of elevated idling error $E_{\rm idle}$ around $\varDelta_{10}=0,{\rm MHz}$ and $\varDelta_{10}=\pm 250,{\rm MHz}$ correspond to the ge-ge and ge-ef frequency-collision conditions between Q0 and Q1.
The white dashed curve in the lower panel indicates the condition under which the idling error satisfies $E_{\rm idle}=0.02\%$.
}
\label{fig:2Q_idling}
\end{figure}

The numerical setup and the corresponding quantum circuit for the time evolution are shown in Fig.~\ref{fig:2Q_idling}(a) and (b).  
The frequency of Q0 was fixed at $\omega_0/(2\pi)=5000\,{\rm MHz}$, while the detuning of Q1 relative to Q0, $\varDelta_{10} := \omega_1 - \omega_0$, was treated as a sweep parameter together with $g$.  
Energy relaxation and dephasing were neglected in this calculation.
In the simulation, Q1 was prepared as a spectator qubit in the maximally mixed state $\hat{I}/2$.  
During a $700\,{\rm ns}$ idling period, an $X$ gate was applied to Q0 at $t=350\,{\rm ns}$ and $t=700\,{\rm ns}$ to dynamically decouple static ZZ interactions.  
The quantum process experienced by Q0 during this idling interval was characterized via quantum process tomography, and the corresponding $\chi$ matrix was extracted.
The $(I,I)$ element of the resulting $\chi$ matrix, $\chi_{II}$, represents the fidelity with respect to the identity operation during idling.  
We therefore define the idling gate error as $E_{\rm idle} = 1 - \chi_{II}$.

Figure~\ref{fig:2Q_idling}(c) shows the numerical results (upper panel) and the fitting results (lower panel) of $E_{\rm idle}$ as functions of $\varDelta_{10}$ and $g$.  
The fitting function used is given by
\begin{align}
&E\sub{fitting}(\varDelta_{10},g)=A\left(\frac{g}{\varDelta_{10}}\right)^{2}\no\\
&\hspace{1cm}+B\left(\frac{g}{\varDelta_{10}-\alpha_{0}}\right)^{2}
+B\left(\frac{g}{\varDelta_{10}+\alpha_{1}}\right)^{2},
\label{eq:1}
\end{align}
where the fitted parameters are $A=1.856$ and $B=2.024$.
The numerical results show that the spectator-induced idling error decreases as the qubit detuning $\varDelta_{10}$ shifts away from the ge-ge and ge-ef collision conditions.  
By solving Eq.~(\ref{eq:1}) for $g$, we obtain the functional dependence of the coupling strength, $g = g(E_{\rm idle}, \varDelta_{10})$, expressed in terms of the allowed idling error and qubit detuning.
In this study, we preliminarily estimated the impact of $g$ on the yield of a large-scale integrated qubit architecture and selected an acceptable idling error $E_{\rm idle}=0.02\%$.  
Then, the corresponding value of $g$ was adopted throughout this study.  
The white dashed curve in Fig.~\ref{fig:2Q_idling}(c) indicates the relationship between $g$ and $\varDelta_{10}$ for $E_{\rm idle}=0.02\%$.

\subsection{Frequency-collision conditions for two qubits under siZZle driving}\label{sec:parametric}

\begin{figure}[t]
\centering
\includegraphics[width=8cm]{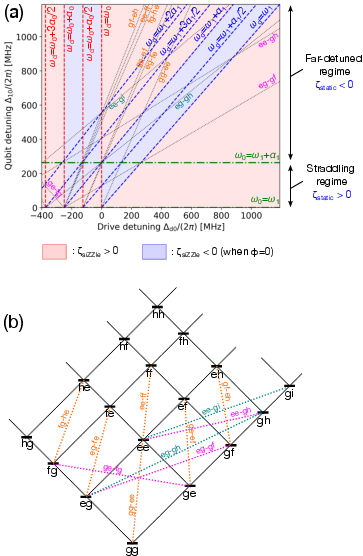}
\caption{
(a) Principal frequency-collision conditions in the parameter space spanned by the siZZle sweep parameters $\varDelta_{\rm d0}$ and $\varDelta_{10}$.
The lightly shaded red and blue regions indicate the parameter regimes in which, for a siZZle phase difference $\phi=0$, the siZZle-induced ZZ interaction satisfies $\zeta_{\rm siZZle} > 0$ (red) and $\zeta_{\rm siZZle} < 0$ (blue), respectively.
(b) Energy-level structure of the two transmons and the corresponding indirect transition processes.
The labels (g, e, f, h) in the left and right slots denote the energy eigenstates of Q0 and Q1, respectively.
}
\label{fig:parametric}
\end{figure}

As shown in Fig.~\ref{fig:1}(c), the fidelity of the siZZle-CZ gate depends strongly on the two sweep parameters, the drive detuning $\varDelta_{\rm d0}$ and the qubit detuning $\varDelta_{10}$.  
In particular, the low-fidelity regions appearing as linear features originate from frequency-collision conditions either between the resonance frequencies of the two qubits or between the qubit resonance frequencies and the siZZle drive frequency.
Figure~\ref{fig:parametric}(a) summarizes the principal frequency-collision conditions in the parameter space spanned by $\varDelta_{\rm d0}$ and $\varDelta_{10}$.  
The green dash-dotted lines indicate the ge-ge and ge-ef collision conditions between the two qubits.  
The boundary defined by $\omega_0 = \omega_1 + \alpha_1$ separates the straddling regime (below the line) from the far-detuned regime (above the line).
The red and blue wavy curves denote collisions between the siZZle drive frequency and the resonance frequencies of Q0 and Q1, respectively.  
The five curves corresponding to the conditions $\omega\sub{d}=\omega_i$, $\omega_i+\alpha_i/2$, $\omega_i+\alpha_i$, $\omega_i+3\alpha_i/2$, and $\omega_i+2\alpha_i$ ($i=0,1$) represent the cases in which the drive frequency matches the g-e, g-f (two-photon), e-f, e-h (two-photon), f-h transition frequencies, respectively.  
In addition, the black dotted lines indicate conditions where the drive induces higher-order indirect transitions distinct from the direct resonances listed above.  
Figure~\ref{fig:parametric}(b) illustrates the energy-level structure of the two transmons and the corresponding indirect transition processes.


The signs of the static ZZ interaction, $\zeta_{\rm static}$, and siZZle-induced ZZ interaction, $\zeta_{\rm siZZle}$, also depend on the relationship between the two sweep parameters $\varDelta_{\rm d0}$ and $\varDelta_{10}$.  
As shown in Fig.~\ref{fig:parametric}(a), in the straddling regime (below the green dash-dotted line defined by $\omega_0 = \omega_1 + \alpha_1$), the static ZZ interaction satisfies $\zeta_{\rm static} > 0$, whereas in the far-detuned regime (above the line), $\zeta_{\rm static} < 0$.
The lightly shaded red and blue regions indicate parameter regimes in which the sign of the siZZle-induced ZZ interaction depends on the relative phase $\phi$ of the siZZle drive.  
For $\phi = 0$, the siZZle-induced ZZ interaction satisfies $\zeta_{\rm siZZle} > 0$ in the red region and $\zeta_{\rm siZZle} < 0$ in the blue region, while the sign is reversed for $\phi = \pi$.
Therefore, to generate a large effective ZZ interaction via siZZle driving, the relative phase $\phi = 0$ or $\pi$ must be selected such that $\zeta_{\rm static}$ and $\zeta_{\rm siZZle}$ interfere constructively under the given frequency conditions.

\subsection{Evaluation method for frequency collisions}\label{sec:collision}

Here we describe the criteria and evaluation procedure for frequency collisions employed in the yield analysis of the integrated qubit chip presented in Sec.~\ref{sec:yield}.  
The overall procedure is as follows. 
For a given qubit frequency allocation on the lattice, we evaluate the depolarizing error accumulated during the siZZle-CZ gate time for each qubit node and for each edge on which a siZZle-CZ gate is applied. 
A frequency collision occurs if the total depolarizing error at any node or edge exceeds a predetermined target error threshold $E_{\star}$.
The following section presents the details of the evaluation procedure.

\subsubsection{Assumed error mechanisms}\label{sec:error}

We consider the following error contributions in the yield evaluation.

(i) {\it Single qubit decoherence error}: 
This error arises from the energy relaxation and dephasing of each qubit in the average siZZle-CZ gate duration, which is assumed to be $700\,\mathrm{ns}$; it is individually evaluated for each qubit node.  
Using numerical simulations, we computed the time evolution of an isolated single qubit idling for $700\,\mathrm{ns}$ with $T_1 = T_2^* = 600\,\mathrm{\mu s}$, and the resulting infidelity with respect to the identity operation, $7.7\times10^{-4}$, is defined as the single-qubit decoherence error.  
In the present numerical study, $T_1$ and $T_2^*$ are set to identical values for all qubits; therefore, the single-qubit decoherence error is uniform across all the qubit nodes.

(ii) {\it Static-spectator error}: 
This error accounts for unwanted interactions between a given qubit and its NN and NNN spectator qubits during the $700\,\mathrm{ns}$ siZZle-CZ gate interval.  
It is evaluated for each qubit node as well as for edges on which the siZZle-CZ gate is applied.
We assume that the target qubit has a resonance frequency of $\omega_0$ and anharmonicity $\alpha_0$, and a given spectator qubit has a resonance frequency $\omega\sub{s}$ and anharmonicity of $\alpha\sub{s}$.  
We define the detuning between the target and spectator qubits as $\varDelta_{\rm s0} := \omega\sub{s} - \omega_0$ and denote their coupling strength by $g$.  
The static-spectator error induced by this spectator qubit on the target qubit is given by $E_{\mathrm{fitting}}(\varDelta_{\rm s0}, g)$, which is defined in Eq.~(\ref{eq:1}).
For the NNN spectator qubits, the effective coupling strength $g_{0,2}$ between the target and NNN spectator qubits is given by
\begin{align}
g_{0,2} = \sum_{i}\frac{g_{0,{\rm c}i}g_{{\rm c}i,2}}{2}
\left(
\frac{1}{\omega_0-\omega_{{\rm c}i}}+\frac{1}{\omega_2-\omega_{{\rm c}i}}
\right),
\label{eq:g}
\end{align}
where ${\rm c}i$ denotes the label of an intermediate qubit located between the target and NNN spectator qubits.  
Here, $\omega_{{\rm c}i}$ is the frequency of the intermediate qubit ${\rm c}i$, and $g_{0,{\rm c}i}$ and $g_{{\rm c}i,2}$ represent the coupling strengths between the target and intermediate ${\rm c}i$ qubit and between the intermediate ${\rm c}i$ and NNN spectator qubits, respectively.  
In a square lattice, two such intermediate qubits exist; therefore, the sum in Eq.~(\ref{eq:g}) runs over all the intermediate qubits ${\rm c}i$.

\begin{figure*}[t]		
\centering
\includegraphics[width=18cm]{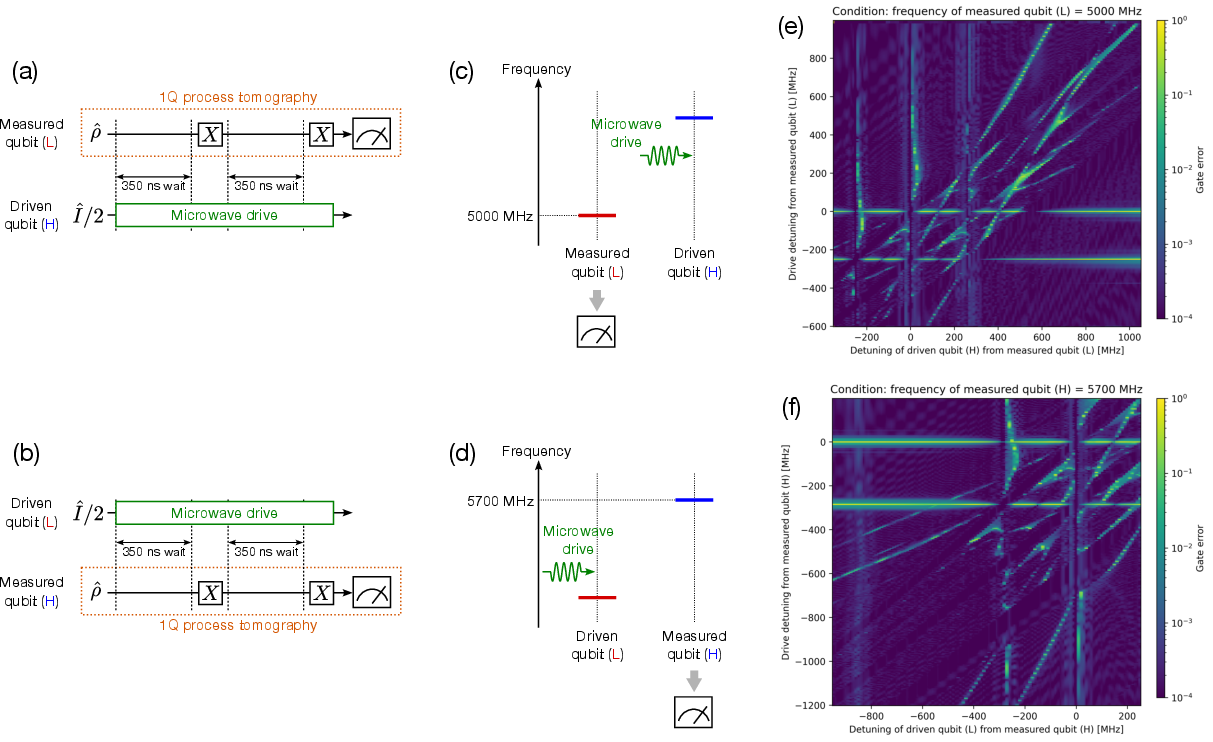}
\caption{
(a), (b) Quantum circuits used to numerically evaluate the driven-spectator error.
The resonance frequencies of the measured qubit are set as (a) $5000\,\mathrm{MHz}$ and (b) $5700\,\mathrm{MHz}$.
(c), (d) Frequency configurations of the measured qubit, driven qubit, and microwave drive corresponding to the conditions in (a) and (b), respectively.
The frequencies of the driven qubit and the microwave drive are swept.
(e), (f) Numerical results of the driven-spectator error as functions of the detuning of the driven qubit and the microwave drive from the measured qubit, corresponding to the conditions in (a) and (b), respectively.
}
\label{fig:dynamical_error}
\end{figure*}

(iii) {\it Driven-spectator error}: 
This error arises when NN and NNN spectator qubits are driven by the siZZle microwave tone and induce unwanted effects on the target qubit; it is evaluated for each qubit node.  
The driven-spectator error depends on the target qubit frequency, the frequency of the driven spectator qubit, the siZZle drive frequency, the siZZle pulse amplitude, and the pulse duration.
In the present evaluation, the target qubit frequency is fixed to either $5000\,\mathrm{MHz}$ (``L'', low-frequency qubit) or $5700\,\mathrm{MHz}$ (``H'', high-frequency qubit), as shown in Fig.~\ref{fig:3}(a), (b).  
The siZZle pulse duration is fixed to the average siZZle-CZ gate time of $700\,\mathrm{ns}$.  
The siZZle pulse amplitude is determined for each spectator-qubit frequency and drive frequency according to the procedure described in Sec.~\ref{sec:sizzle_params}.
The driven-spectator error as a function of the spectator-qubit frequency and siZZle drive frequency is evaluated numerically using QuTiP, as shown in Fig.~\ref{fig:dynamical_error}.  
Figures~\ref{fig:dynamical_error}(a) and (b) show the quantum circuits used for the numerical evaluation when the target qubit is assigned to ``L'' and ``H'', respectively.  
Figures~\ref{fig:dynamical_error}(c) and (d) illustrate the relative ordering of the target-qubit frequency, spectator-qubit frequency, and siZZle drive frequency under each condition.  
Figures~\ref{fig:dynamical_error}(e) and (f) present the numerical results of the driven-spectator error obtained by sweeping the spectator-qubit frequency and siZZle drive frequency.
The driven-spectator error is enhanced under conditions such as static frequency collisions between the target and spectator qubits, CR-like conditions, and other indirect transition resonances.  
For NNN spectator qubits, the effective coupling strength given by Eq.~(\ref{eq:g}) is used in the evaluation.

(iv) {\it Individual siZZle-CZ gate error}: 
This term represents the intrinsic gate error of the siZZle-CZ operation itself and is evaluated for each edge on which the siZZle-CZ gate is applied.  
The error is obtained from the results shown in Fig.~\ref{fig:1}(c) as a function of the detuning between the two qubits involved in the siZZle-CZ gate and the siZZle drive frequency.  
By construction, this error excludes spectator-induced errors acting on the two qubits participating in the siZZle-CZ gate.
The siZZle drive frequency is selected, as detailed in Sec.~\ref{sec:how_to_det_drive_freq}, to maximize the siZZle-CZ gate fidelity while ensuring that the driven-spectator error experienced by the surrounding spectator qubits does not exceed the target error threshold $E^\star$.

\subsubsection{Procedure for frequency-collision evaluation}\label{sec:tejun}

\begin{figure*}[t]		
\centering
\includegraphics[width=18cm]{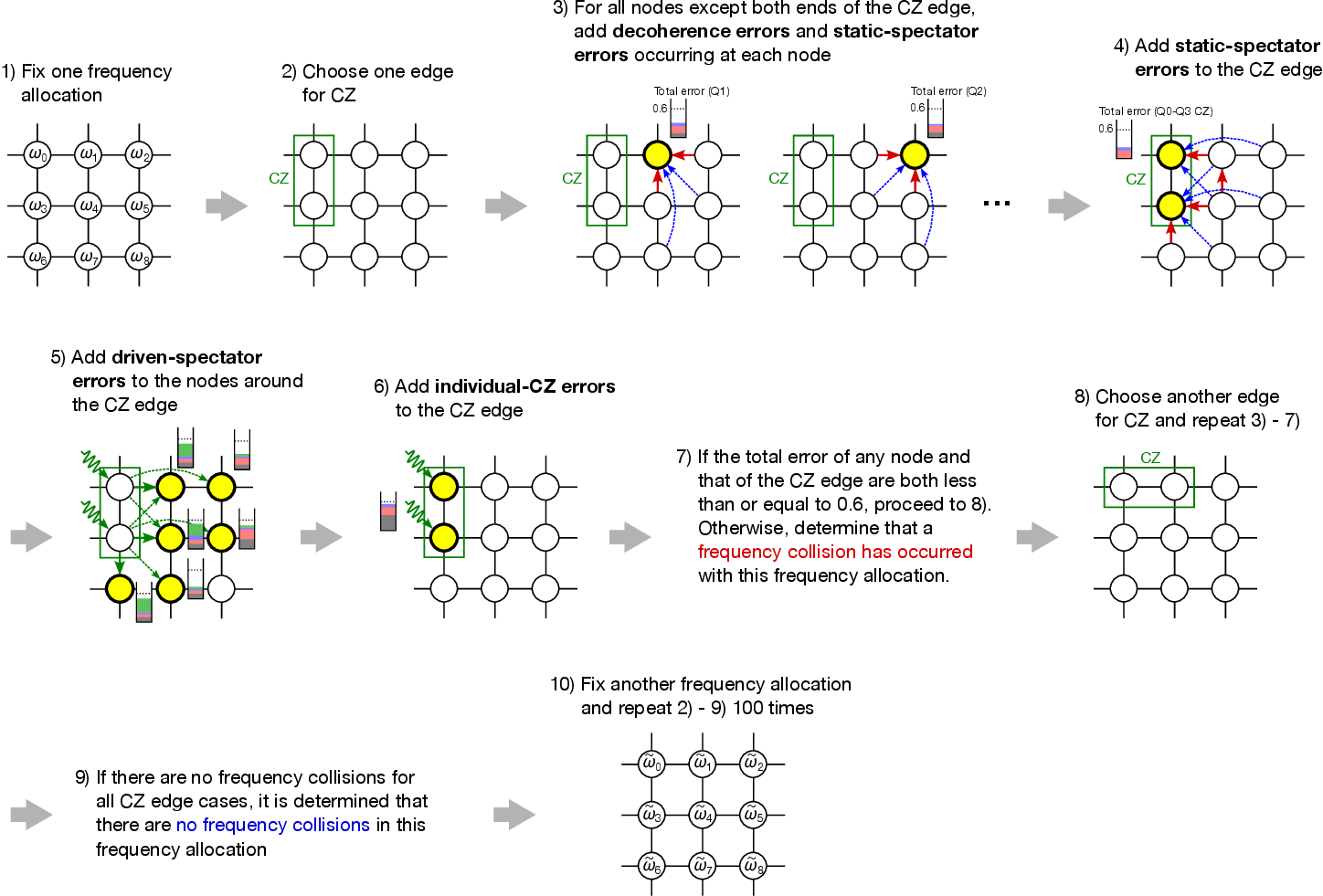}
\caption{
Evaluation flow for the zero-collision yield of an integrated qubit lattice with a given frequency design.
In steps 1) and 10), when fixing the frequency assignment, stochastic frequency offsets are added to the design frequencies to model fabrication-induced frequency dispersion.
In steps 5) and 6), the siZZle pulse drive frequency is chosen according to the procedure described in Fig.~\ref{fig:how_to_determine}, such that the driven-spectator error on nodes adjacent to the CZ edge is minimized while the CZ gate error is simultaneously minimized.
Under this selected condition, the corresponding driven-spectator error and CZ gate error are added to the total error budget.
Although this figure illustrates the case of a square lattice, the same procedure applies to the heavy-hexagonal lattice.
}
\label{fig:collision}
\end{figure*}

Figure~\ref{fig:collision} illustrates the procedure for calculating the zero-collision yield of an integrated qubit lattice with a given frequency design and lattice structure.

\def\labelenumi{\arabic{enumi})}
\begin{enumerate}

\item  
A resonance frequency is assigned to each node of the qubit lattice.
To model fabrication-induced frequency dispersion, a random frequency offset is added to each design frequency, sampled from a normal distribution with a specified standard deviation.

\item One edge is selected from all edges in the lattice and designated as the edge on which the siZZle-CZ gate is applied (hereafter referred to as the ``CZ edge'').

\item For all qubit nodes except the two endpoints of the CZ edge (hereafter referred to as the ``CZ nodes''), the single-qubit decoherence error and static-spectator error are evaluated and summed.  
In the evaluation of the static-spectator error, NN and NNN qubits excluding the CZ nodes are treated as spectator qubits.

\item The static-spectator error for the two CZ nodes is evaluated and added to the total error of the CZ edge.  
Here as well, NN and NNN qubits excluding the CZ nodes are considered as spectator qubits.

\item For all qubit nodes other than the CZ nodes, the driven-spectator error is evaluated and added to each node's total error.  
In this evaluation, driven spectator qubits at NN and NNN positions are considered.  
The siZZle-CZ drive frequency used here is determined according to the procedure described in Sec.~\ref{sec:how_to_det_drive_freq}.

\item The individual siZZle-CZ gate error is added to the total error of the CZ edge.  
The siZZle drive frequency is identical to that used in step 5).

\item Then, we verify whether the total error at the CZ edge and at all the non-CZ qubit nodes remains below the target error threshold $E_\star$ (set to $0.6\%$ in this study).  
If the error at any CZ edge or qubit node exceeds $E_\star$, then the frequency assignment generated in step 1) is deemed to exhibit a frequency collision.

\item If no frequency collision is detected in step 7), then a different edge is selected as the CZ edge, and steps 3)--7) are repeated.  
If a frequency collision occurs for any choice of CZ edge, then the frequency assignment generated in step 1) is deemed to exhibit a frequency collision.

\item If no frequency collision occurs for any possible choice of CZ edge in the assumed lattice, then the frequency assignment generated in step 1) is deemed to be collision-free.

\item To compute the zero-collision yield, new random frequency assignments are generated probabilistically, and steps 2)--9) are repeated a sufficient number of times (100 iterations in this study).  
The zero-collision yield under the given design conditions is defined as the fraction of frequency assignments for which no collision occurs.

\end{enumerate}

\subsubsection{Determination of the siZZle drive frequency}\label{sec:how_to_det_drive_freq}

When the resonance frequencies of the two qubits involved in a siZZle-CZ gate are fixed, the drive frequency that minimizes the siZZle-CZ gate error can be selected simply by following the numerical results shown in Fig.~\ref{fig:1}(c). 
In practice, however, spectator qubits exist in the vicinity of the two qubits participating in the siZZle-CZ interaction. 
As discussed in Sec.~\ref{sec:error}, these spectator qubits can experience driven-spectator errors, as illustrated in Fig.~\ref{fig:dynamical_error}(e) and (f). 
Therefore, to avoid frequency collisions, the drive frequency must be selected such that the driven-spectator errors on all the spectator qubits remain sufficiently small and that the siZZle-CZ gate error is simultaneously minimized.

\begin{figure*}[t]		
\centering
\includegraphics[width=18cm]{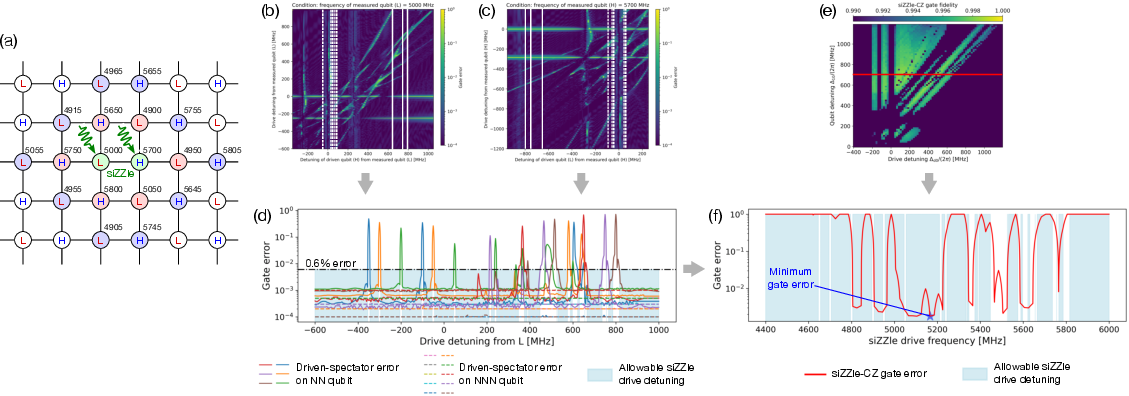}
\caption{
Procedure for determining the siZZle pulse drive frequency. 
(a) Qubits driven by the siZZle pulse (green), NN qubits of the driven qubit (red), and NNN qubits (blue). 
The numbers shown at the upper-right of each qubit indicate the qubit resonance frequencies (in MHz) for an example frequency allocation. 
The resonance frequencies of the qubits labeled L and H are assumed to be around 5000 and 5700\,MHz, respectively. 
(b), (c) Numerical results for the driven-spectator error reproduced from Fig.~\ref{fig:dynamical_error}(e) and (f). 
The white solid and dashed lines indicate the conditions corresponding to the resonance frequencies of the NN and NNN qubits in panel (a), respectively. 
(d) Driven-spectator errors for each NN qubit (solid lines) and NNN qubit (dashed lines) in the example of panel (a), plotted as a function of the siZZle drive frequency. 
The dash-dotted line indicates a gate-error threshold of 0.6\%. 
The region where all driven-spectator errors remain below this threshold is highlighted in light blue and defined as the allowable siZZle drive-frequency range. 
(e) Numerical results for the siZZle-CZ gate fidelity reproduced from Fig.~\ref{fig:1}(c). 
The red solid line corresponds to the qubit detuning of 700\,MHz in the example shown in panel (a). 
(f) siZZle-CZ gate error (defined as one minus the gate fidelity shown in panel (e), red solid line) plotted together with the allowable siZZle drive-frequency range obtained in panel (d) (light-blue region). 
The frequency that yields the minimum gate error within this allowable range is used as the optimal siZZle drive frequency for the example in panel (a).
}
\label{fig:how_to_determine}
\end{figure*}

A concrete example of the procedure used to determine the siZZle drive frequency is shown in Fig.~\ref{fig:how_to_determine}. 
Figure~\ref{fig:how_to_determine}(a) illustrates the layout on a square lattice, indicating the two qubits participating in the siZZle-CZ interaction (green), NN spectator qubits (red), and NNN spectator qubits (blue). 
The resonance frequencies of the qubits labeled ``L'' and ``H'' are assigned around 5000 and 5700\,MHz, respectively, and the numbers shown at the upper-right of each qubit node indicate the example of the resonance frequency of that qubit in units of MHz.
To identify the range of siZZle drive frequencies for which the driven-spectator errors on all NN and NNN spectator qubits remain sufficiently small, we use the numerical results for the driven-spectator error shown in Fig.~\ref{fig:dynamical_error}(e) and (f). 
Figures~\ref{fig:how_to_determine}(b) and (c) reproduce these numerical results, where the resonance frequencies of the NN and NNN spectator qubits in the example of panel (a) are indicated by solid and dashed lines, respectively. 
Figure~\ref{fig:how_to_determine}(d) shows the driven-spectator errors for these NN and NNN spectator qubits as a function of the detuning of the siZZle drive frequency from the resonance frequency of the ``L''-labeled qubit involved in the siZZle-CZ gate (5000\,MHz in this example). 
The dash-dotted line indicates the target error threshold (here $E_\star=0.6\%$). 
The frequency region in which none of the driven-spectator errors exceeds this threshold (highlighted in light blue) is taken as the allowable range for the siZZle drive frequency.

Next, to determine the drive frequency within this allowable range that minimizes the siZZle-CZ gate error, we use the numerical results for the siZZle-CZ gate fidelity shown in Fig.~\ref{fig:1}(c). 
Figure~\ref{fig:how_to_determine}(e) reproduces these results, where the qubit detuning between the two qubits involved in the siZZle-CZ gate in the example of panel (a) is indicated by a red solid line. 
Figure~\ref{fig:how_to_determine}(f) plots the siZZle-CZ gate error along this red line together with the allowable frequency range for the siZZle drive frequency. 
The blue star marks the drive frequency that yields the minimum siZZle-CZ gate error within the allowable region. 
In the frequency-collision evaluation described in Sec.~\ref{sec:tejun}, this optimization procedure is performed for each candidate frequency allocation and for each CZ edge to determine the optimal siZZle drive frequency.

\section*{Data availability}
All data supporting the findings of this study are available within the article. 
Additional data are available from the corresponding author upon reasonable request.

\begin{acknowledgments}
This research was supported by 
JST PRESTO (Grant No.~JPMJPR23F2), 
JST COI-NEXT (Grant No.~JPMJPF2014), 
JST Moonshot R\&D (Grants No.~JPMJMS2067), 
and MEXT Q-LEAP (Grant No.~JPMXS0118068682) 
\end{acknowledgments}

\section*{Author contributions}
K.O.~conceived the project, developed the theoretical framework, and performed the numerical simulations. 
K.O.~analyzed the results and wrote the manuscript. 
All authors contributed to discussions of the results and approved the final manuscript.

\section*{Competing Interests}
The authors declare no financial or non-financial competing interests.

\clearpage

\appendix





\end{spacing}

\nocite{*}
\bibliography{ref}

\end{document}